\documentclass{article}
\usepackage{amsmath,amssymb,amsthm,graphicx,relsize,tocloft,soul,xcolor,booktabs,longtable,textcomp}
\usepackage[parfill]{parskip}    
\usepackage{float}
\usepackage{subcaption}
\usepackage{hyperref}

\usepackage{natbib}

\usepackage{etoolbox}
\patchcmd{\thebibliography}{\section*{\refname}}{}{}{}

\begin{document}

\title{Feature-Based Likelihood Ratios for Forensic Science: Combining Neural Networks with Bayesian Probability Calculus}
\author{P. Vergeer \\ Netherlands Forensic Institute \\ \href{mailto:p.vergeer@nfi.nl}{p.vergeer@nfi.nl}} 
\date{\today}
\maketitle
\textbf{Keywords:} likelihood ratio, neural network, feature-based, score-based, two level model, PLDA, TLM

\begin{abstract}
\noindent In forensic science, when crime-scene evidence (CSE) and suspect-related evidence (SRE) is present, it is custom to report on the value of this evidence in the form of a likelihood ratio (LR). The LR can be calculated as the probability of CSE given SRE (and that the CSE was generated by the suspect) divided by the probability of CSE given that it was generated by a randomly selected person from an alternative culprit population (for which we have a relevant dataset). In forensic science, this is known as a feature-based  LR and intuitively the LR contrasts ‘similarity’ by ‘typicality’. 
Since it is generally a problem for feature-based LRs to find appropriate models for the data, one either resorts to score-based LRs – modeling probabilities of a scalar comparison function of the difference between CSE and SRE – or to adjusting the feature-based output post-hoc to well-calibrated output. Either way, the above definition of the LR is broken and interpretation of the LR as similarity between CSE and SRE divided by typicality of CSE is destroyed.

Here, we report on progress in obtaining instantly well-performing feature-based LRs using gradient descent in combination with Bayesian probability theory – integrating out uncertainty of nuisance parameters -  to train as a neural network a two-level model, the main LR-model in forensic science for describing distributions of continuous data. 
For a dataset of laser-ablation inductively-coupled-plasma mass-spectrometry measurements on glass fragments from forensic casework, we show that our best model on validation data yields much better calibrated feature-based LRs on the test set when compared to state-of-the-art feature-based LR-systems trained on the same type of data, and that it improves a factor 4.5 on average on the main performance metric (Cllr).  

For LRs interpretable in terms of ‘similarity’ contrasting ‘typicality’ this is a major advancement. However, a state-of-the-art LR-system still performs a factor of 1.5 better on Cllr for this data. We also present future plans to close this remaining gap. In order to facilitate collaboration, we have put relevant code on Github.

\end{abstract}

\section{Introduction}
\label{Intro}

In forensic science, we are often confronted with the question of 'origin of source' of some material found at the crime scene. The question then is how strong the link is between this crime-scene material and a source related to a suspect (the source can be the suspect itself). In order to answer this question, the forensic scientist takes measurements of the crime-scene object, we will denote these measurements as the crime-scene evidence (CSE), and measurements from the known source or suspect, we will denote these measurements as suspect-related evidence (SRE). It is common to evaluate this collective evidence in the light of two hypotheses,

\begin{itemize}
    \item[] $H_1:$ \textit{The CSE was deposited by the identified source}
    \item[] $H_2:$ \textit{The CSE was deposited by a randomly selected source in the alternative source population. The SRE is unrelated.}
\end{itemize}

and report on the value of evidence in the form of a likelihood ratio (LR)\cite{AitkenPositionStatement, AitkenStatisticsBook, BergerPositionStatement, AitkenTheUseOfStatistics, MorrisonPositionStatement, MorrisonIso, RobertsonVignauxBergerBook}.

Advancing the paradigm shift to more data-driven methods in forensics \cite{SaksParadigmShiftScience, MorrisonParadigmShiftDataScience}, over the past 20 years or so numerical methods to calculate likelihood ratios (these numerical methods are further referred to as LR-systems) have been published for many forensic evidence types, including speaker comparison\cite{MorrisonForensicSpeechBook, GonzalezSpeakerVerification, VandervloedDataSelection}, face comparison\cite{MinagliaFaceComparisonDifferentQuality}, firearm toolmark comparison\cite{BaikerToolmarkGun}, fingerprint comparison\cite{AbrahamFingerPrintReview, AlberinkFingerPrint, LiFingerPrintLR, LeegwaterFingerPrintLR}, gasoline-to-fire-debris comparison\cite{VergeerGasolineFireDebris}, handwriting comparison\cite{BozzaHandwritingLRs, HeplerHandwritingLRs, JohnsonHandwringLR},  MDMA-profiling\cite{BolckMDMAScoreFeatureBasedLRs, BolckMDMTLM}, footwear-mark comparison\cite{EvettFootwareLRs}, and glass-fragment comparison\cite{EsVergeerGlassLRs, AkmeemaneAlmirallGlass}. In this literature on forensic LR-systems, broadly speaking, two classes of LR-systems are developed: so-called feature-based (FB) and score-based LR-systems\cite{BolckMDMAScoreFeatureBasedLRs, HeplerHandwritingLRs, VergeerRankingTheStars, NeumannDefenceAgainstTwo}. Feature-based LR-systems model the probability of CSE and SRE under both hypotheses, and adhere to the definition of the likelihood ratio in terms of probabilities of the evidence under both hypotheses. A feature-based LR ($LR_{FB}$) is:

\begin{align}\label{eq:1}
LR_{FB} := \frac{P(CSE, SRE | H_1)}{P(CSE, SRE | H_2)}.
\end{align}

Usually, the denominator is written as a product of two independent probabilities\cite{BolckMDMTLM, AitkenTLM, AitkenTraceEvidenceMultivariateData}, $$P(CSE, SRE | H_2)  = P(CSE | H_2) \times P(SRE | H_2),$$ and $P(SRE | H_2) = P(SRE | H_1) = P(SRE) $ so that Eq. \ref{eq:1} becomes,

\begin{align}\label{eq:2}
LR_{FB} := \frac{P(CSE | SRE, H_1)}{P(CSE | H_2)}.
\end{align}

Intuitively, the numerator of Eq. \ref{eq:2} is known as representing 'similarity' between CSE and SRE, while the denominator represents 'typicality' of CSE when generated by the alternative source population.

Score-based LRs lose this interpretation of 'similarity' divided by 'typicality' of CSE\cite{MorrisonScoreBasedTypicality, NeumannDefenceAgainstTwo, MorrisonTakingAccountOfTypicality}, as they reduce the information \cite{VergeerRankingTheStars, BoonstraReconciling} in CSE and SRE to a comparison score based on the difference between CSE and SRE measurement values. Score-based LRs do not adhere to the definition of the LR in Eq. \ref{eq:1} since they do not model the probability of the joint evidence. By construction, the score ($S$) becomes the new evidence, and $S$ is calibrated to a score-based LR by defining a mapping function $S \rightarrow LR(S)$. Such a mapping function does not have to be based on modeling the numerator and denominator score-distributions explicitly; one can also find a mapping function directly, e.g. by logistic regression\cite{MorrisonTutorialLogisticRegression} or Pool Adjacent Violators calibration\cite{BrummerApplication-independent}.

From a practical perspective, score-based LRs are relatively easy to obtain, since they concern a one-dimensional variable and the mapping function is relatively easy to find\cite{VergeerRankingTheStars}. On the other hand, FB LRs are hard to obtain, since data is usually multidimensional and probability models that fit the data well are hard to define. For continuous data, one usually resorts to assumptions of multivariate normality\cite{AitkenTLM, BolckMDMTLM}. The 'workhorse' feature-based LR-model for continuous data is known as the 'Two Level Model'(TLM) and assumes multivariate normality at two levels: between sources and between measurements within sources\cite{Lindley1977, AitkenTLM, BolckMDMTLM}, and an extension to multivariate KDE methods for the between sources distribution exists\cite{AitkenTLM, BolckMDMTLM} and is also used\cite{EsVergeerGlassLRs, AkmeemaneAlmirallGlass, BolckMDMTLM}. In the speaker recognition community, the 'normalnormal' TLM is also known as probabilistic linear discriminant analysis\cite{brummerPLDA, GlemberPLDA, CumaniPLDA, SizovPLDA} (PLDA).

The ill-fitting normality assumption in conjunction with estimating an inflated number of parameters (means and covariance matrices) results in a poor fit and ill-calibrated LRs, e.g. for glass data\cite{EsVergeerGlassLRs, RamirezGaussianization, RamosAlmiralWithinUncertainty} or speaker comparison data\cite{RibeiroEffectsGausianization}. The current bottom line is: feature-based LRs as such do not make it even though they have the advantageous 'similarity'' divided by 'typicality' interpretation, since the data is simply too difficult to model directly.

In order to fix this deficiency, LR-output from these systems may be calibrated post-hoc\cite{EsVergeerGlassLRs, VandervloedDataSelection}, resulting in values that are well-calibrated and improve decisions\cite{VergeerRankingTheStars, BoonstraReconciling}. Unfortunately, 1) ill model fit results in information loss, and 2) when mapping the feature-based LR-output to a well-calibrated LR, the post-hoc calibrated LR loses the interpretation in terms of 'similarity' divided by 'typicality'. 

So from a theoretical perspective, feature-based LR-systems are attractive, since when one does find a good model for the data - resulting in well-calibrated FB-LRs directly - one does not lose the typicality information a score-based LR loses and superior performance is expected\cite{VergeerRankingTheStars}. Therefore, recently several attempts have been made to 'bring the data to the model', by Gaussianization of measurements on glass fragments\cite{RamirezGaussianization} and of speaker embeddings from deep neural networks (DNNs)\cite{RibeiroEffectsGausianization}. These works aim to preprocess the random variables that TLM operates on in order to align their distribution with the characteristics of a multivariate Gaussian distribution.

Also attempts have been made to 'bring the model to the data', by trying to improve predictions by using a full-Bayesian approach by integrating out the uncertainty in the covariance matrix estiamtes\cite{RamosAlmiralWithinUncertainty, RamirezGaussianization}, or making between models more flexible\cite{RamosAlmiralWithinUncertainty}.

These studies report performance gains, bringing log-likelihood-ratio cost (Cllr\cite{BrummerApplication-independent}) down from about 1 to 0.15 for laser-ablation inductively-coupled-plasma mass-spectrometry (LA-ICP-MS) measurements on glass fragments\cite{RamirezGaussianization}, and from 0.189 to 0.163 for DNN embeddings for speaker comparison (male-matched condition)\cite{RibeiroEffectsGausianization}. These performance gains are encouraging, and arguably more encouraging is that there is also room for more improvement, since calibration is still poor. For example, the calibration error makes up for 70$\%$ of the quoted Cllr for the glass measurements, while the histograms in \cite{RibeiroEffectsGausianization}, Fig. 13-15, still show dissimilarity compared to their theoretical benchmarks.

On the other hand, the performance gap between Gaussianized feature-based LR systems and post-hoc calibrated feature-based LR-systems is still expected to be large. On comparable data as the measurements in \cite{RamirezGaussianization} (which are on glass data from Bundes Kriminal Ambt casework) a post-hoc calibrated FB LR-system on glass data from the Netherlands Forensic Institute (NFI) reaches a Cllr of 0.021, more than a factor of 7 smaller than the best procedure in \cite{RamirezGaussianization} on the BKA data. For such comparable datasets (both casework, glass, LA-ICP-MS), the difference for the two methods is substantial and there remains a gap to be bridged.

The present study tries to narrow this gap by training the parameters of the TLM as a neural network. It is hypothesized that the main cause of the large calibration error for FB LRs is that the maximum-likelihood (ML) method to estimate the mean and covariance matrices is not suitable for a 'likelihood-ratio objective' when the data distribution deviates from multivariate normal. When training the TLM as a neural network, one can directly optimize on the likelihood-ratio output (by using the Cllr as criterion) and likely obtain different parameter estimates (compared to the ML-estimates) that yield a better-calibrated, and hopefully better performing, LR-system.

Training the TLM as a neural network is not new\cite{RamojiNPLDAFirst, RamojiNPLDAGithub, RamojiNPLDA_CSL, BurgetBrummerNPLDA}. In the speaker-recognition field, it is known as 'neural PLDA' (NPLDA)\cite{RamojiNPLDAFirst, RamojiNPLDAGithub, RamojiNPLDA_CSL}. These authors implement NPLDA as a quadratic scoring layer (giving log likelihood ratios) with as trainable parameters two positive semi-definite matrices (they center there data beforehand so the grand mean is a vector of zeros and does not have to be trained). As criterion they use a differentiable approximation to the normalized detection cost. This criterion measures discrimination and is insensitive to calibration. Burget et al. \cite{BurgetBrummerNPLDA} train the parameters of the TLM (the two covariance matrices and the grand mean) directly and have a computationally efficient algorithm that exploits the redundancy of instances over all pairs. They optimize on Cllr and denote 30$\%$ performance gain over the ML-estimated PLDA baseline. They also implement a neural 'heavy-tailed PLDA' model, which can be interpreted as integrating out covariance parameter uncertainty. This gives a 40$\%$ improvement over the baseline model, but the authors see no way to train this model in a computationally efficient way. The authors do not report the Cllr-values themselves, only relative improvements.

\subsection{Contribution of this work}
We see several ways to improve the current state-of-the-art of NPLDA training.
\newline
First, we will use the singular-value representation of covariance matrices to disentangle the trainable parameters into two distinct categories: variances and projection matrices).
This allows us to de-correlate the omnibus log$_{10}$ likelihood ratio (LLR) into a sum of independent LLRs, which makes our implementation more legible than the current state-of-the-art, and also allows us to manipulate a specific parameter category, fully leveraging the flexibility of neural net training by implementing or relaxing constraints on parameters.
\newline
Second, we will leverage the statistical-model interpretation of the Neural TLM (NTLM) layer by first selecting the best model using a point estimate for the diagonal within-variance matrix (as usual), but when we predict on the test set we will first integrate out the within-variance parameter uncertainty determined on the validation set. We note that this 'integrate-out procedure' has been tried before, but is was applied during training\cite{BurgetBrummerNPLDA}, while we only use it for out-of-sample prediction, doing gradient descent using the original TLM and integrating out the variance post-hoc.
\newline
Third, the TLM assumes a common within-variance for all sources. We will relax this assumption and fit a distribution for the within-variance among sources.
\newline
Last, in forensic science, most of the time we have 'small' data (i.e. max 1000 sources with max 10 repetitions). Therefore, we also propose a training pipeline that is specifically designed for the 'small data' regime.

The outline of the paper is as follows.
The methods section introduces the glass dataset, our NTLM formulation, and the experimental/computational setup.
Then the usual results section follows, and we will end with a discussion, future plans and conclusion.

Core functions of our torch neural net implementation can be found on \href{https://github.com/vergep/NeuralNetLikelihoodRatios}{Github}: https://github.com/vergep/NeuralNetLikelihoodRatios.
\section{Methods}
\subsection{The glass dataset}

\subsubsection{Instrumental}
The glass dataset comprises LA-ICP-MS measurements obtained in NFI casework from 1,053 fragments originating from reference glass panels collected between 2020 and 2024. Measurements were performed using an NWR193 193-nm excimer laser-ablation system, inductively coupled to a plasma mass spectrometer (IC-MS; iCAP RQ+). The laser was operated at a laser power of 30 mW with a spot size of 60 $\mu$m. An FGS02 glass standard was used for external calibration. Isotopes measured were Li7, Na23, Mg24, Al27, Si28, K39, Ca42, Ti49, Mn55, Fe57, Rb85, Sr88, Zr90, Ba137, La139, Ce140, Nd146, Hf180, Pb208. For each panel, 6 repetitions were measured. For our 'small-data' study, we selected randomly 2 repetitions for each panel.

\subsubsection{Preprocessing}
Signals were stabilized over measurements by dividing isotope signals to the Si28 isotope signal, and subsequently the logarithm was taken. After splitting the 1053 sources in train, validate and test datasets of equal size (351 sources each), data were standardized by subtracting the grand mean of the train data and division by the standard deviation of the train data.

Tables \ref{tab:MeanWithinCov} and \ref{tab:BetweenCov} show the mean covariance matrix for repetions and the covariance matrix for the glass panel means respectively.

\begin{table}[htbp]
\centering
\resizebox{\textwidth}{!}{%
\begin{tabular}{lrrrrrrrrrrrrrrrrrr}
 & Al27 & Ba137 & Ca42 & Ce140 & Fe57 & Hf180 & K39 & La139 & Li7 & Mg24 & Mn55 & Na23 & Nd146 & Pb208 & Rb85 & Sr88 & Ti49 & Zr90 \\
\hline
Al27 & 2.78E-04 & 1.24E-04 & 1.84E-04 & 9.09E-05 & 7.49E-05 & 3.25E-04 & 3.85E-05 & 2.45E-04 & 1.75E-05 & 9.49E-05 & 5.66E-05 & 3.08E-05 & 2.34E-04 & 2.71E-05 & 3.26E-05 & 1.85E-04 & 1.91E-04 & 3.27E-04 \\
Ba137 & 1.24E-04 & 1.52E-04 & 9.64E-05 & 6.07E-05 & 5.44E-05 & 1.39E-04 & 2.96E-05 & 1.27E-04 & 2.38E-05 & 5.60E-05 & 4.36E-05 & 2.51E-05 & 1.21E-04 & 4.17E-05 & 3.20E-05 & 1.07E-04 & 9.69E-05 & 1.31E-04 \\
Ca42 & 1.84E-04 & 9.64E-05 & 1.52E-04 & 7.58E-05 & 6.18E-05 & 2.19E-04 & 3.35E-05 & 1.79E-04 & 2.26E-05 & 7.43E-05 & 5.02E-05 & 2.88E-05 & 1.66E-04 & 2.24E-05 & 2.86E-05 & 1.37E-04 & 1.40E-04 & 2.22E-04 \\
Ce140 & 9.09E-05 & 6.07E-05 & 7.58E-05 & 1.01E-04 & 4.55E-05 & 1.28E-04 & 2.48E-05 & 9.71E-05 & 3.02E-05 & 4.86E-05 & 4.25E-05 & 2.69E-05 & 9.30E-05 & 3.28E-05 & 3.20E-05 & 7.52E-05 & 8.11E-05 & 1.18E-04 \\
Fe57 & 7.49E-05 & 5.44E-05 & 6.18E-05 & 4.55E-05 & 9.11E-05 & 9.46E-05 & 2.89E-05 & 7.76E-05 & 3.78E-05 & 4.63E-05 & 4.38E-05 & 2.97E-05 & 6.47E-05 & 3.17E-05 & 3.05E-05 & 6.29E-05 & 6.45E-05 & 9.22E-05 \\
Hf180 & 3.25E-04 & 1.39E-04 & 2.19E-04 & 1.28E-04 & 9.46E-05 & 9.94E-04 & 5.09E-05 & 3.20E-04 & 3.24E-05 & 1.13E-04 & 7.34E-05 & 4.41E-05 & 3.23E-04 & 3.77E-05 & 6.09E-05 & 2.20E-04 & 2.39E-04 & 6.95E-04 \\
K39 & 3.85E-05 & 2.96E-05 & 3.35E-05 & 2.48E-05 & 2.89E-05 & 5.09E-05 & 7.49E-05 & 3.57E-05 & 2.92E-05 & 2.52E-05 & 2.77E-05 & 3.22E-05 & 3.35E-05 & 4.17E-05 & 4.42E-05 & 2.99E-05 & 3.77E-05 & 4.79E-05 \\
La139 & 2.45E-04 & 1.27E-04 & 1.79E-04 & 9.71E-05 & 7.76E-05 & 3.20E-04 & 3.57E-05 & 3.16E-04 & 2.34E-05 & 9.60E-05 & 6.19E-05 & 3.39E-05 & 2.31E-04 & 3.75E-05 & 3.51E-05 & 1.81E-04 & 1.87E-04 & 3.08E-04 \\
Li7 & 1.75E-05 & 2.38E-05 & 2.26E-05 & 3.02E-05 & 3.78E-05 & 3.24E-05 & 2.92E-05 & 2.34E-05 & 2.74E-04 & 2.60E-05 & 2.92E-05 & 3.55E-05 & 4.79E-05 & 2.94E-05 & 1.93E-05 & 2.39E-05 & 2.95E-05 & 2.28E-05 \\
Mg24 & 9.49E-05 & 5.60E-05 & 7.43E-05 & 4.86E-05 & 4.63E-05 & 1.13E-04 & 2.52E-05 & 9.60E-05 & 2.60E-05 & 6.22E-05 & 3.67E-05 & 2.76E-05 & 9.01E-05 & 2.05E-05 & 2.67E-05 & 7.70E-05 & 7.66E-05 & 1.16E-04 \\
Mn55 & 5.66E-05 & 4.36E-05 & 5.02E-05 & 4.25E-05 & 4.38E-05 & 7.34E-05 & 2.77E-05 & 6.19E-05 & 2.92E-05 & 3.67E-05 & 6.56E-05 & 2.80E-05 & 5.18E-05 & 2.77E-05 & 2.44E-05 & 5.14E-05 & 5.26E-05 & 7.04E-05 \\
Na23 & 3.08E-05 & 2.51E-05 & 2.88E-05 & 2.69E-05 & 2.97E-05 & 4.41E-05 & 3.22E-05 & 3.39E-05 & 3.55E-05 & 2.76E-05 & 2.80E-05 & 4.42E-05 & 3.55E-05 & 3.62E-05 & 3.45E-05 & 2.95E-05 & 3.33E-05 & 4.06E-05 \\
Nd146 & 2.34E-04 & 1.21E-04 & 1.66E-04 & 9.30E-05 & 6.47E-05 & 3.23E-04 & 3.35E-05 & 2.31E-04 & 4.79E-05 & 9.01E-05 & 5.18E-05 & 3.55E-05 & 5.72E-04 & 3.31E-05 & 3.50E-05 & 1.67E-04 & 1.82E-04 & 3.03E-04 \\
Pb208 & 2.71E-05 & 4.17E-05 & 2.24E-05 & 3.28E-05 & 3.17E-05 & 3.77E-05 & 4.17E-05 & 3.75E-05 & 2.94E-05 & 2.05E-05 & 2.77E-05 & 3.62E-05 & 3.31E-05 & 2.66E-04 & 5.48E-05 & 2.37E-05 & 3.06E-05 & 3.85E-05 \\
Rb85 & 3.26E-05 & 3.20E-05 & 2.86E-05 & 3.20E-05 & 3.05E-05 & 6.09E-05 & 4.42E-05 & 3.51E-05 & 1.93E-05 & 2.67E-05 & 2.44E-05 & 3.45E-05 & 3.50E-05 & 5.48E-05 & 2.03E-04 & 2.67E-05 & 4.17E-05 & 5.04E-05 \\
Sr88 & 1.85E-04 & 1.07E-04 & 1.37E-04 & 7.52E-05 & 6.29E-05 & 2.20E-04 & 2.99E-05 & 1.81E-04 & 2.39E-05 & 7.70E-05 & 5.14E-05 & 2.95E-05 & 1.67E-04 & 2.37E-05 & 2.67E-05 & 1.59E-04 & 1.38E-04 & 2.22E-04 \\
Ti49 & 1.91E-04 & 9.69E-05 & 1.40E-04 & 8.11E-05 & 6.45E-05 & 2.39E-04 & 3.77E-05 & 1.87E-04 & 2.95E-05 & 7.66E-05 & 5.26E-05 & 3.33E-05 & 1.82E-04 & 3.06E-05 & 4.17E-05 & 1.38E-04 & 1.99E-04 & 2.39E-04 \\
Zr90 & 3.27E-04 & 1.31E-04 & 2.22E-04 & 1.18E-04 & 9.22E-05 & 6.95E-04 & 4.79E-05 & 3.08E-04 & 2.28E-05 & 1.16E-04 & 7.04E-05 & 4.06E-05 & 3.03E-04 & 3.85E-05 & 5.04E-05 & 2.22E-04 & 2.39E-04 & 7.42E-04 \\
\hline
\end{tabular}%
}
\caption{Mean covariance matrix for repetitions}
\label{tab:MeanWithinCov}
\end{table}

\begin{table}[htbp]
\centering
\resizebox{\textwidth}{!}{%
\begin{tabular}{lrrrrrrrrrrrrrrrrrr}
 & Al27 & Ba137 & Ca42 & Ce140 & Fe57 & Hf180 & K39 & La139 & Li7 & Mg24 & Mn55 & Na23 & Nd146 & Pb208 & Rb85 & Sr88 & Ti49 & Zr90 \\
\hline
Al27 & 4.11E-02 & 2.83E-02 & -1.81E-03 & 2.88E-02 & 6.83E-03 & 4.81E-03 & 3.79E-02 & 2.34E-02 & 8.17E-03 & 3.81E-03 & 1.23E-02 & 6.38E-04 & 2.28E-02 & 1.16E-02 & 3.99E-02 & 5.88E-03 & 1.78E-02 & 4.60E-03 \\
Ba137 & 2.83E-02 & 8.12E-02 & -1.87E-03 & 8.77E-03 & 5.52E-04 & 5.98E-03 & 7.57E-02 & 6.32E-03 & 6.70E-03 & -5.16E-03 & 2.78E-03 & 6.46E-05 & 2.53E-03 & 2.48E-02 & 6.48E-02 & 2.05E-02 & 9.31E-03 & 6.04E-03 \\
Ca42 & -1.81E-03 & -1.87E-03 & 1.87E-03 & -1.15E-03 & -1.85E-03 & -7.43E-04 & -2.36E-03 & -5.30E-04 & -8.13E-05 & -3.41E-03 & -5.29E-04 & -2.52E-05 & -4.72E-04 & 8.25E-04 & -2.79E-03 & 7.42E-04 & 4.43E-05 & -7.00E-04 \\
Ce140 & 2.88E-02 & 8.77E-03 & -1.15E-03 & 1.03E-01 & 1.75E-02 & 1.30E-02 & -1.19E-03 & 7.80E-02 & 9.60E-03 & 6.69E-03 & 5.01E-02 & 9.12E-04 & 8.29E-02 & 1.74E-02 & 1.09E-02 & -4.07E-04 & 4.03E-02 & 1.36E-02 \\
Fe57 & 6.83E-03 & 5.52E-04 & -1.85E-03 & 1.75E-02 & 1.02E-01 & 5.30E-03 & 8.47E-03 & 1.12E-02 & -1.79E-03 & -4.74E-04 & 1.30E-02 & 9.96E-05 & 9.93E-03 & -1.68E-04 & 1.24E-02 & -3.12E-03 & 2.06E-02 & 4.78E-03 \\
Hf180 & 4.81E-03 & 5.98E-03 & -7.43E-04 & 1.30E-02 & 5.30E-03 & 3.30E-02 & 3.46E-03 & 1.20E-02 & -6.52E-03 & 1.67E-03 & -7.19E-03 & 2.45E-04 & 1.21E-02 & 2.52E-03 & 3.65E-03 & 1.10E-03 & 1.63E-02 & 3.41E-02 \\
K39 & 3.79E-02 & 7.57E-02 & -2.36E-03 & -1.19E-03 & 8.47E-03 & 3.46E-03 & 1.41E-01 & -6.57E-03 & 1.03E-02 & -6.05E-04 & -2.46E-02 & -5.10E-04 & -1.10E-02 & 2.45E-02 & 1.28E-01 & 1.68E-02 & -2.27E-03 & 2.59E-03 \\
La139 & 2.34E-02 & 6.32E-03 & -5.30E-04 & 7.80E-02 & 1.12E-02 & 1.20E-02 & -6.57E-03 & 6.85E-02 & 5.58E-03 & 7.03E-03 & 4.93E-02 & 9.07E-04 & 7.29E-02 & 1.68E-02 & 2.22E-03 & 1.48E-03 & 3.44E-02 & 1.27E-02 \\
Li7 & 8.17E-03 & 6.70E-03 & -8.13E-05 & 9.60E-03 & -1.79E-03 & -6.52E-03 & 1.03E-02 & 5.58E-03 & 2.49E-02 & 1.60E-03 & 1.88E-02 & 2.22E-04 & 6.78E-03 & 7.82E-03 & 1.74E-02 & -7.28E-03 & 1.94E-03 & -6.78E-03 \\
Mg24 & 3.81E-03 & -5.16E-03 & -3.41E-03 & 6.69E-03 & -4.74E-04 & 1.67E-03 & -6.05E-04 & 7.03E-03 & 1.60E-03 & 2.74E-02 & 1.03E-02 & 4.88E-04 & 7.77E-03 & -4.63E-03 & 1.51E-03 & -3.64E-03 & 1.35E-03 & 1.80E-03 \\
Mn55 & 1.23E-02 & 2.78E-03 & -5.29E-04 & 5.01E-02 & 1.30E-02 & -7.19E-03 & -2.46E-02 & 4.93E-02 & 1.88E-02 & 1.03E-02 & 1.06E-01 & 8.10E-04 & 5.45E-02 & 2.19E-02 & -1.57E-02 & 2.37E-04 & 2.72E-02 & -6.68E-03 \\
Na23 & 6.38E-04 & 6.46E-05 & -2.52E-05 & 9.12E-04 & 9.96E-05 & 2.45E-04 & -5.10E-04 & 9.07E-04 & 2.22E-04 & 4.88E-04 & 8.10E-04 & 2.16E-04 & 1.04E-03 & -1.74E-04 & -3.03E-04 & -2.94E-05 & 5.20E-04 & 2.77E-04 \\
Nd146 & 2.28E-02 & 2.53E-03 & -4.72E-04 & 8.29E-02 & 9.93E-03 & 1.21E-02 & -1.10E-02 & 7.29E-02 & 6.78E-03 & 7.77E-03 & 5.45E-02 & 1.04E-03 & 7.95E-02 & 1.65E-02 & 5.75E-05 & -7.05E-04 & 3.53E-02 & 1.28E-02 \\
Pb208 & 1.16E-02 & 2.48E-02 & 8.25E-04 & 1.74E-02 & -1.68E-04 & 2.52E-03 & 2.45E-02 & 1.68E-02 & 7.82E-03 & -4.63E-03 & 2.19E-02 & -1.74E-04 & 1.65E-02 & 8.78E-02 & 2.33E-02 & 5.61E-03 & 1.33E-02 & 2.60E-03 \\
Rb85 & 3.99E-02 & 6.48E-02 & -2.79E-03 & 1.09E-02 & 1.24E-02 & 3.65E-03 & 1.28E-01 & 2.22E-03 & 1.74E-02 & 1.51E-03 & -1.57E-02 & -3.03E-04 & 5.75E-05 & 2.33E-02 & 1.32E-01 & 1.26E-03 & 1.75E-03 & 2.60E-03 \\
Sr88 & 5.88E-03 & 2.05E-02 & 7.42E-04 & -4.07E-04 & -3.12E-03 & 1.10E-03 & 1.68E-02 & 1.48E-03 & -7.28E-03 & -3.64E-03 & 2.37E-04 & -2.94E-05 & -7.05E-04 & 5.61E-03 & 1.26E-03 & 3.99E-02 & -7.95E-04 & 1.44E-03 \\
Ti49 & 1.78E-02 & 9.31E-03 & 4.43E-05 & 4.03E-02 & 2.06E-02 & 1.63E-02 & -2.27E-03 & 3.44E-02 & 1.94E-03 & 1.35E-03 & 2.72E-02 & 5.20E-04 & 3.53E-02 & 1.33E-02 & 1.75E-03 & -7.95E-04 & 4.17E-02 & 1.70E-02 \\
Zr90 & 4.60E-03 & 6.04E-03 & -7.00E-04 & 1.36E-02 & 4.78E-03 & 3.41E-02 & 2.59E-03 & 1.27E-02 & -6.78E-03 & 1.80E-03 & -6.68E-03 & 2.77E-04 & 1.28E-02 & 2.60E-03 & 2.60E-03 & 1.44E-03 & 1.70E-02 & 3.56E-02 \\
\hline
\end{tabular}%
}
\caption{Covariance matrix of glass panel means}
\label{tab:BetweenCov}
\end{table}

\begin{figure}[H]
    \centering
    \includegraphics[width=1\textwidth]{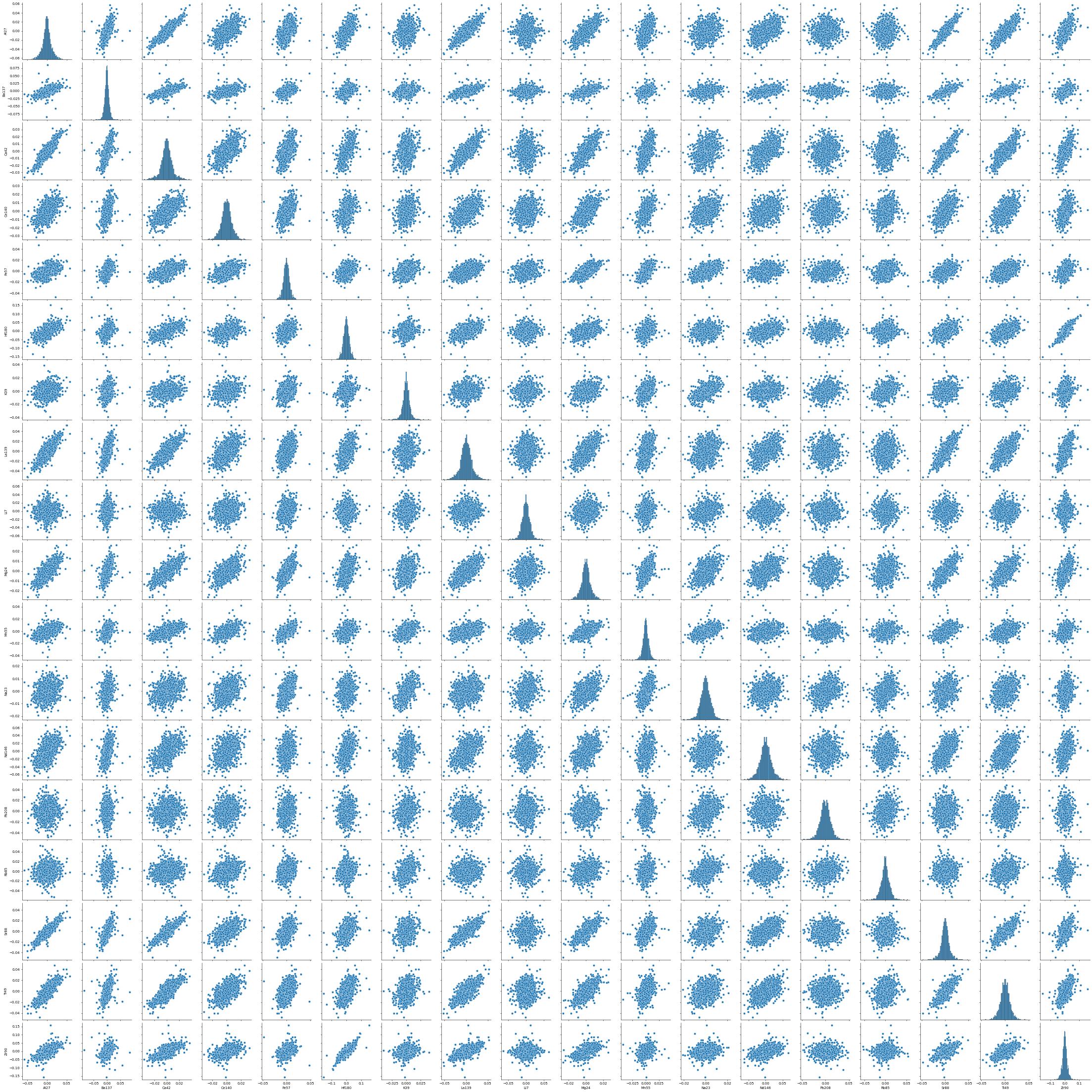}
    \caption{Bi-variate scatterplots of repetitions minus their means}
    \label{fig:WithinBivariateScatter}
\end{figure}

\begin{figure}[H]
    \centering
    \includegraphics[width=1\textwidth]{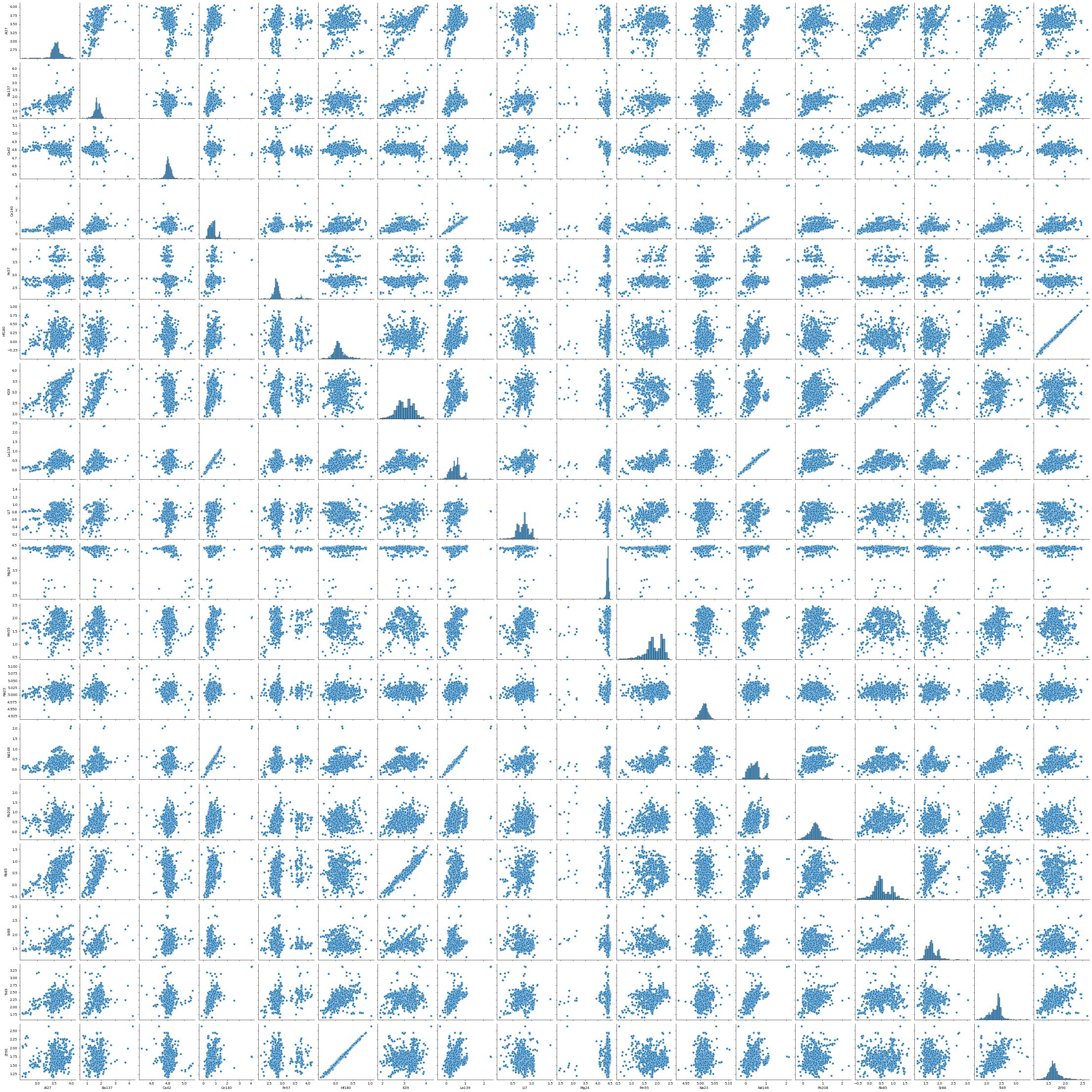}
    \caption{Bi-variate scatterplots of glass panel means.}
    \label{fig:BetweenBivariateScatter}
\end{figure}

Figure \ref{fig:WithinBivariateScatter} shows bi-variate scatterplots of the 6 repetitions with their means subtracted, and Fig. \ref{fig:BetweenBivariateScatter} shows bi-variate scatterplots of the panel means.

These plots have 18 by 18 panels, showing log ppm signals for combinations of two of the 18 elements in the off-diagonal panels and histograms of the log ppm signals for an element on the diagonal. These plots can be used to eyeball the goodness of normality assumptions for the Tlm. The scatterplots of the repetitions (Fig. \ref{fig:WithinBivariateScatter}) show that the distributions are somewhat close to multivariate normal whereas the scatterplots of the glass panel means (Fig. \ref{fig:BetweenBivariateScatter}) show larger deviations from multivariate normality.

\subsection{Small-data validation protocol}
A regular deep learning data protocol is to have three fixed datasets, train, validate and test. The train dataset is used for the gradient descent feedback signal. The validation dataset is used for calculating the value of a metric during training and to select the model with best performance. In this way several models can be trained and the best one (on the validation metric) is selected. Finally, the unseen test dataset is used to get an unbiased estimate of the performance.\cite{WatsonDeepLearningWithPython}

Since we regard our dataset size as 'small' for the purpose of neural net training, we have used a modified version of a small-data protocol in \cite{WatsonDeepLearningWithPython} known as 'iterated K-fold validation with shuffling'. These authors recommend applying K-fold cross-validation, applied P times. This procedure first splits off the test set. Subsequently, the train/validate data is split in K folds, and iteratively one fold is used as validation dataset and the rest as train dataset. This K-fold protocol is done P times starting with a random shuffle of the train/validate data. The average validation-score is used to select the best model.

However, we found that using this procedure, the performance of the final model is strongly dependent on the particular randomly assigned test data. In order to accommodate this variance, we split our data into train/validation/test datasets, and do this P times starting with a random shuffle of the complete dataset. We study not only average performance but also its spread on validation and test data. 

Note that at the moment we are not interested in bringing a model to production. We have a scientific interest: how well does a model do in this small-data setting, on average performance and its spread. Therefore, we 'reuse' our data by partitioning it multiple times and do not use an extra 'overall' hold-out set.

Arbitrarily, we set P to 21 (since a regular experiment then still runs overnight).
Experiments were performed using a predecessor of the LiR package\cite{lir}.

\subsection{Train/validate pipeline}
The train/validate pipeline (programmed using Pytorch) is as follows:
\begin{enumerate}
\item Split sources into train, validate and test, 351, 351, 351 sources each.
\item Pair all instances per split, and label the pairs same or different source.
\item Prepare a 'dataloader' that provides same- and different-source pairs to the model.
The same-source batch size was set at 50; the dataloader calculates the best accompanying different-source batch size (8750 pairs in this case).
Left-over same-source comparisons (351 MOD 50 = 1) are spread evenly over the other batches. This prevents a noisy update at the end of an epoch.
\item Train the model using the train and validate dataloaders.
The training specifications were:
\begin{itemize}
\item loss: Cllr, both for training, validation and test
\item optimizer: AdamW, learning rate: 0.001
\item train for 2000 epochs and save the model with the lowest validation loss
\end{itemize}
\end{enumerate}

\subsection{TLM formulation as a neural network}
\begin{figure}[H]
    \centering
    \includegraphics[width=1\textwidth]{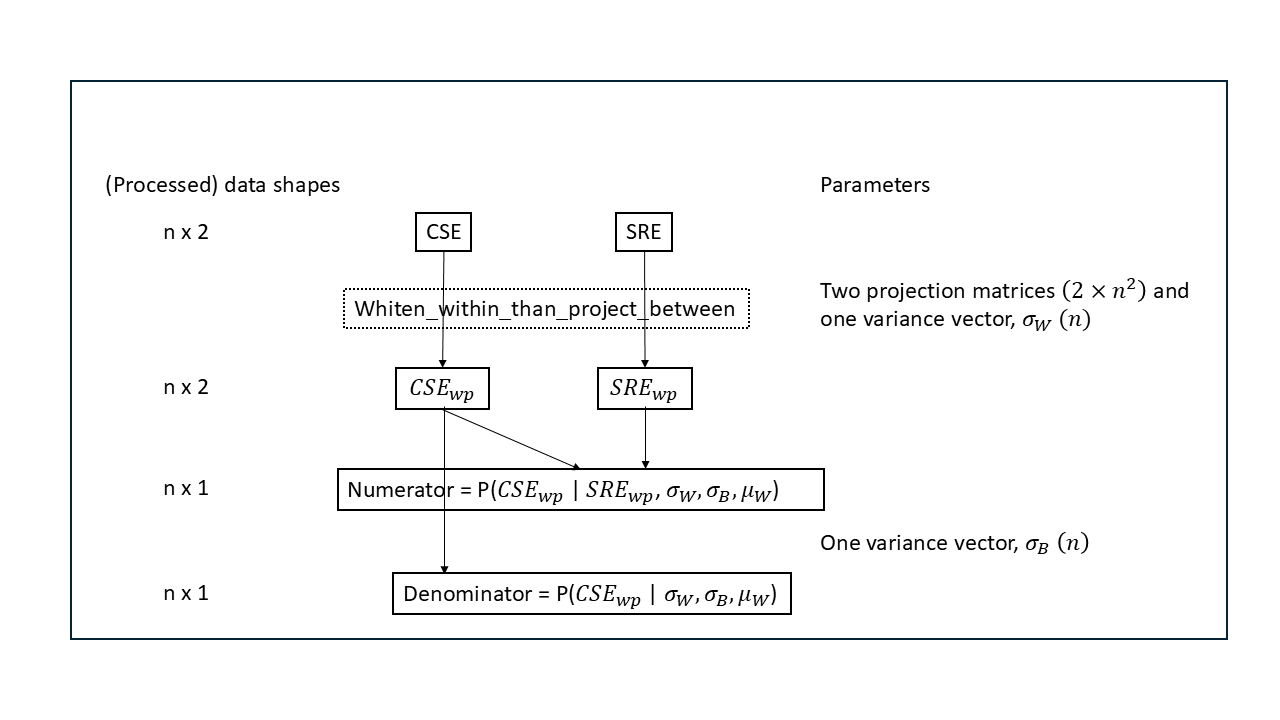}
    \caption{Sketch of NTLM-model with two projection matrices and one within-variance vector.}
    \label{fig:model_sketch}
\end{figure}

A sketch of our model pipeline (with two projection matrices) is given in Fig. \ref{fig:model_sketch}.

Row-vectors of CSE and SRE are the input for the NTLM-model.We will further denote one such a row-vector as an instance.
Instances are linearly decorrelated in both the within and between space by first whithening the data in the within space, and then projecting this whitened data on the principal axes for the between space:

Let $Q_w$ and $Q_b$ be two orthonormal matrices whose column vectors represent axes. Let $\sigma_w$ be the within-variance vector. Then, whitened-than-projected data $d_{wp}$ for instance $d$ is obtained by:

\begin{align}\label{eq:3}
d_{wp} = d . Q_w .diag(1/\sqrt(sigma_w)) . Q_b,
\end{align}

where $.$ denotes a dot product and 'diag' denotes a matrix with non-zero entries on the diagonal only. The space of all orthonormal matrices was spanned by using Q from a QR-decomposition (torch.linalg.qr) of a square matrix of random numbers (which are the trainable parameters). The order: whiten within, project between (and not whithen between, project within) was chosen since the within data is the closest to multivariate normality (see Fig. \ref{fig:WithinBivariateScatter}).

Per axis, the numerator and denominator probabilities were calculated as posterior and prior predictive distribution probabilities using a Gaussian Mixture Model (GMM) for the between distribution. For a NormalNormal TLM, the GMM has only one component, with mean $\mu_B$ and variance $\sigma_B$. Posterior and prior predictive distributions can be derived by standard Bayesian probability calculus for normal distributions\cite{Bolstad}. Note that for GMM operations $\sigma_w$ is a diagonal matrix with 1's on the diagonal due to the whitening operation.

After the decorrelation operation, the processed data is treated as independent for all axes, i.e. the omnibus LLR is the sum over axes of LLRs per axis.

\subsubsection{From two projection matrices to one}
When training the NTLM-model with one projection matrix, the whiten-than-project operation was replaced by one projection operation. GMM operations now use a within- and between-variance vector as trainable parameters.

\subsubsection{From one to two within-variance vectors}
When training the NTLM-model with two within-variance vectors, the GMM operations use a different within-variance vector for the processed (whitened-than-projected or project) CSE as for the processed SRE.

\subsection{Per-axis analysis}
\subsubsection{LLR-histograms per axis}
Since we have decorrelated numerator and denominator vector elements, we can study properties per axis in the decorrelated space. Same-source and different-source LLR-histograms were made and Cllr-values were calculated per axis.

\subsubsection{QQnorm-plots for projected data under $H_1$ and $H_2$ per axis}

QQnorm plots provide a qualitative way to study normality of data. A QQnorm plot shows the data-quantiles versus the quantiles of a standard normal distribution. If the data follows a standard normal distribution, the plot shows an identity line.

If same-source data follows our $H_1$ model, we assume that to a good approximation, same-source data is standard normally distributed after the following operation:
\begin{align}\label{eq:4}
z_{num} = \frac{CSE_{wp} - SRE_{wp}}{\sigma_*^*},
\end{align}

where $\sigma_*^2 = \sigma_W{_{CSE}}^2 + \sigma_W{_{SRE}}^2$ and $z_{num} \sim StandardNormal$.

This formula is exact for Bayesian updating of the mean of a normal distribution with a flat, improper prior for the mean. It is an approximation for our case, since when calculating LRs we use the grand mean and between variance to define the prior distribution for the glass panel mean. Since z-scores are not the primary objective (LRs are) we use this approximation for ease of (vectorized) calculation.

If $CSE$ data follows our $H_2$ model, $CSE$ data is standard normal distributed after the following operation:
\begin{align}\label{eq:5}
z_{den} = \frac{CSE_{wp} - \mu_B}{\sigma_*^*},
\end{align}

where $\sigma_*^2 = \sigma_W{_{CSE}}^2 + \sigma_B^2$ and $z_{den} \sim StandardNormal$.

\subsection{A NTLM that trains on sources instead of pairs}
We also implemented a NTLM version that trains on batches of sources (providing all instances per source for the sources in a batch). In the future, this is expected to facilitate end-to-end training, since then pairs of data (or embeddings) cannot be calculated beforehand. Our 'Source NTLM' was inspired by the ProxyAnchorLoss implementation\cite{KimProxyAnchorLoss}, where source means from all sources in the data are used as auxiliary trainable parameters and instances of a batch are paired with parameter-vectors representing source-means to have good 'data coverage'. 

Since we have few shots per source, we found that using source means as trainable parameters led to instability in the loss while training.  Therefore, we implemented an analogous, but stable, procedure as follows. 
We start from the premise that for small data and doing end-to-end training (which is not done here), the parameter change after one epoch, and accordingly the change in embedding space, is small.

This justifies the following procedure. From a batch of sources, pair all instances that are from the same source. These are the same-source pairs for this batch. Also, add these instances to a stored list of instances provided in prior batches, with their source labels, under a \texttt{torch.no\_grad} clause. Using these stored instances, create all different-source pairs when pairing the batch of instances with the stored instances. Calculate LRs for all pairs, and subsequently the Cllr.

Note that after one epoch, all instances/labels in the complete dataset are stored this way and we achieve the same data-coverage as for ProxyAnchorLoss. Also, after the first epoch the new batches overwrite a part of these instances. For end-to-end training this would come down to updating the stored embeddings by the batch embeddings.

\subsection{Integrating out variance uncertainty}

\subsubsection{Common within variance, integrating out $H_1$ and $H_2$ variance uncertainty}
Assuming unknown mean and variance, and using a conjugate Normal-inverse gamma prior distribution for these parameters, both numerator and denominator probabilities become t-distributed\cite{wikipediaConjugatePrior}. As prior parameter settings for the normal-inverse gamma distribution we used $\mu$ (prior mean, set to 0 ), $\nu$ (prior sample size for mean, set to 0), $\alpha$ (0.5 times the prior number of observation for the standard deviation of the t-distribution, set to 2) and $\beta_{scaled}$ (prior for the scaled mean-squared deviation, as compared to a mean-squared deviation of 1, set to 3). We divided the data by its mean standard deviation in order to be able to set a conservative $\beta$ using this scaled $\beta$-value. Setting $\alpha = 2$ and $\beta = datavariance \times 3$ we obtain a mean prior variance (mean of the inverse gamma distribution) of three times the data variance, and a mode of the inverse gamma distribution that equals the data variance. We used this as a conservative (skewed to larger variances), non-informative prior distribution for the variance. The number of data-points for posterior updates of the parameters was set to 351.

\subsubsection{Varying within variance, integrating out this uncertainty}
When integrating out the within-variance parameter uncertainty assuming varying within-variance over sources, we defined a prior within-variance distribution by finding parameter values for an inverse gamma distribution. Parameter values were obtained using validation data processed by the best-model for this particular run. Within-variance parameter uncertainty was integrated out numerically by sampling (n = 2000) from this prior distribution, and replacing the data within variance vector by a sampled within variance vector. This was done n times and averages of the prior predictive of the joint and the prior predictive of SRE times the prior predictive of CSE were calculated and the Bayes factor is the ratio of the two, see Eq. \ref{eq:6}.

\begin{align}\label{eq:6}
\frac{P(CSE | SRE)}{P(CSE)} = \frac{P(CSE, SRE)}{P(SRE)P(CSE)} = \frac{\int P(CSE, SRE | \sigma_w)\pi(\sigma_w| \alpha, \beta)d\sigma_w}{\int P(SRE | \sigma_w)\pi(\sigma_w| \alpha, \beta)d\sigma_w\int P(CSE | \sigma_w)\pi(\sigma_w| \alpha, \beta)d\sigma_w},
\end{align}

where $\pi(\sigma_w| \alpha, \beta)$ is the inverse-gamma prior distribution.

Parameter estimates were obtained by expressing the dependence on $\alpha$ and $\beta$ of the sums-of-squares (SSQs) for the validation data (processed by the selected model) as an hiearchical two-level model. For a single source, $\sigma_w$ is distributed according to the inverse gamma model, and one SSQ given $\sigma_w$ is then $\chi$-squared distributed. This lead to a closed form solution writing SSQ as a function of the inverse gamma parameters: 

Under the normal model, $SSQ/\sigma^2\sim\chi^2_{n-1}$. With the prior $\sigma^2\sim\operatorname{InvGamma}(\alpha,\beta)$, it follows that $2\beta/\sigma^2\sim\chi^2_{2\alpha}$. Consequently,

\begin{align}\label{eq:7}
 \frac{SSQ}{2\beta} = \frac{\chi^2_{n-1}/2}{\chi^2_{2\alpha}/2} \sim \operatorname{BetaPrime}\left(\frac{n-1}{2},\alpha\right),
\end{align}
since the ratio of two independent Gamma variables with common rate parameter follows a beta-prime distribution.\cite{SiegristBetaPrime}

Maximum-likelihood parameter estimates were obtained by initializing alpha = 4 and beta = $\frac{mean(SSQs)}{(n-1)(alpha-1)}$, with n the number of repetitions per source (2). We used the 'L-BFGS-B' method in Scipy to minimize the log likelihood of the data.

Maximum a posteriori (MAP) estimates were obtained by setting an exponential prior distribution for $\beta_{map}$ with rate parameter $0.5 \times \beta_{ml}$. This promotes smaller values of $\beta$ resulting in a bias towards larger $\sigma_w$, which generally leads to conservative LLR-values (closer to 0).

Rarely ($\leq 0.001$ of cases) the algorithm did not converge. In these cases, the procedure was retried initializing alpha $\rightarrow$ alpha + 0.5 and redoing this until a convergent solution was reached.

Finally, if the estimated $\alpha$ is smaller than a value of 2, the inverse gamma distribution has infinite variance itself. We consider this result nonphysical for a within-variance distribution, so for these cases we clamped the $\alpha$ parameter to 2 + 1e-6 and kept the mode of the ML-estimated distribution by setting $\beta_{map}$ to

\begin{align}\label{eq:8}
\beta_{map} = \frac{3 \times \beta_{ml}}{\alpha_{ml} + 1 + 1e-6}
\end{align}

\section{Results}

\begin{figure}[H]
    \centering
    \includegraphics[width=1\textwidth]{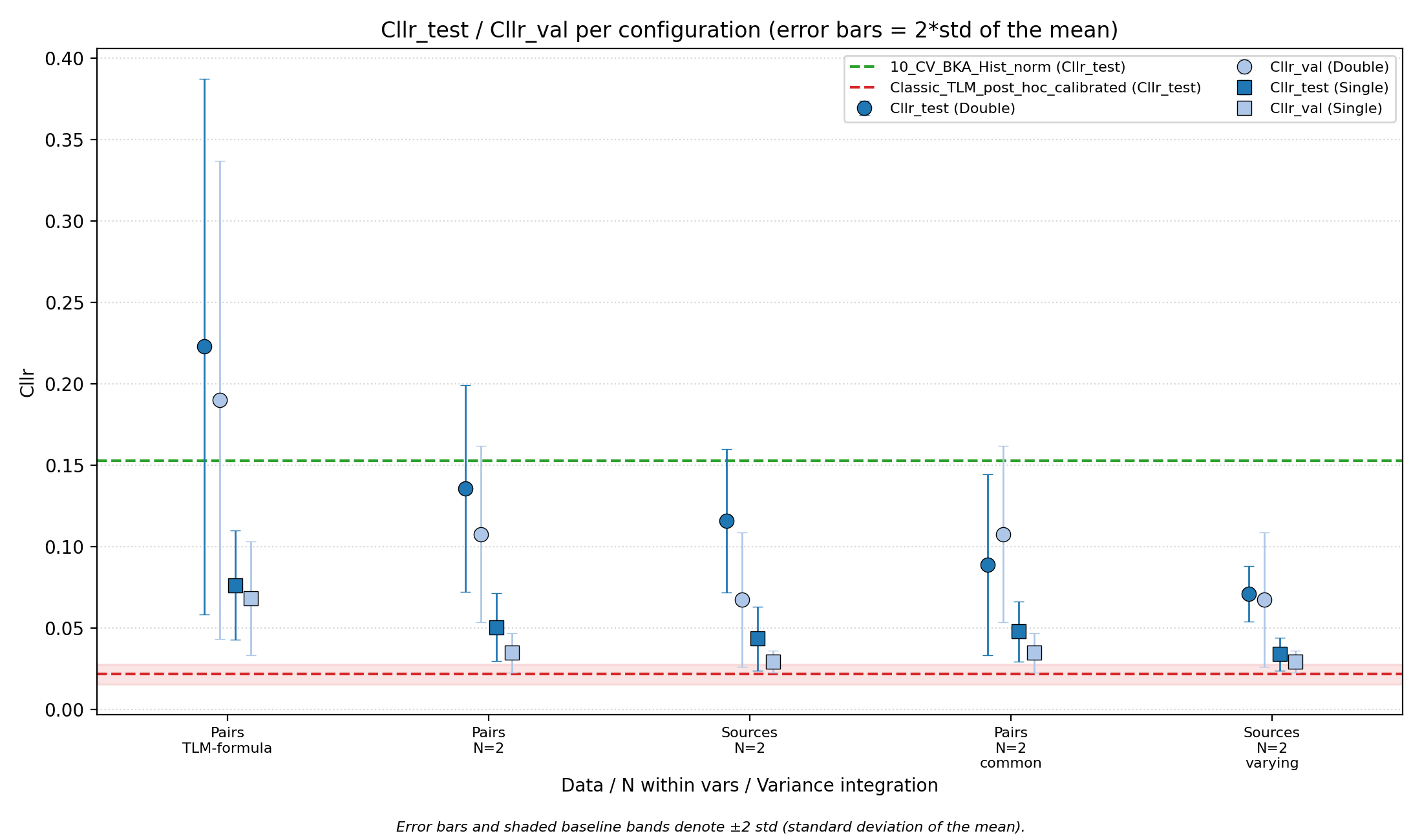}
    \caption{Results for different TLM-NN implementations, compared with benchmarks (horizontal lines). Light blue: validation data results. Dark blue: test data results.}
    \label{fig:grand_summary}
\end{figure}

Figure \ref{fig:grand_summary} gives an overview of the performance of the model types. It pots Cllr (with 2 standard deviation error bars for the means, derived fromt the 21 runs) for the several model types. The horizontal dashed lines denote benchmarks values. The green dashed line is the best model from \cite{RamirezGaussianization}. The red dashed line (red shaded area is two standard deviations from the mean) denotes the NFI state-of-the art model (a post-hoc calibrated TLM LR-system) for this data. Light blue denotes validation data performance, dark blue test data performance. Circles denote models with two projection matrices, and squares denote models with one projection matrix. Independent of further model type, models with one projection matrix outperform models with two projection matrices. The left model ("TLM-formula") encodes the original TLM model, with two covariance matrices and the same within-variance vector for each instance in a pair. The 'N=2' models relax the 'same within variance' constraint and allow the 'CSE' instances of a pair to have a different within-variance vector than the 'SRE' instances of a pair. Comparing the 'TLM-formula' model to the 'Pairs, N=2' model, it can be seen that relaxing this constraint improves performance.

So-called 'Pairs' models were trained with batches of same-source and different-source pairs, whereas 'Sources' models were trained with batches of sources (with all instances per source in one batch). For the 'Sources' models, pairing was done inside the model. Same-source pairs were generated by pairing instances within sources for this batch. Different-source pairs were generated by pairing all instances within a batch with all different-source instances in the complete dataset (which were stored inside the model). 'Sources' models perform slightly better than 'Pairs' models, as can be seen by comparing 'Pairs, N=2' to 'Sources, N=2'.

The 'common' and 'varying' models integrate out the uncertainty of the within variance for the predicted LLRs for the test data. The model 'common' assumes that all sources have the same within variance vector (or vectors for N=2 models), the model 'varying' assumes that sources have different within variances for which a distribution is fitted post-hoc based on the validation data and the fitted TLM-version. As can be seen by comparing 'Pairs, N=2' to the 'Pairs, N=2, common' condition, the latter two models perform slightly better on the test data (for validation data the performance is equal since the 'integrate-out' procedure was only applied to the test data). 

The best-performing NN-model on the right integrates out the variance uncertainty assuming varying within-variance over sources. Not only does this increase performance, it also stabilizes it, showing smaller error-bars for the means than the other models per 'Double/Single' type. Moreover, the 'validate-to-test' performance loss is less.

Overall, the standard TLM NN has has mean Cllr on test data of 0.22 (2 std of mean = 0.16). Several ways to improve on this model are shown. The best NN-model has a mean Cllr on test-data of 0.034 (2 std of mean = 0,01). Compared to the feature-based 'BKA' benchmark that has a Cllr of 0.15\cite{RamirezGaussianization}, this is an improvement of a factor of 4.4. Compared to the 'NFI-state-of-the-art post-coc calibrated TLM' benchmark that has a Cllr of 0.022 (2 std of mean = 0.006), its performance is a factor of 1.5 worse.

Further in is this section, we will show some interesting results for some of the model types, and for some of the 21 seeds. We focus on general trends and exemplify them by showing results from the seeds that gave the median performance on the validation data.

\begin{figure}[H]
    \centering
    \includegraphics[width=1\textwidth]{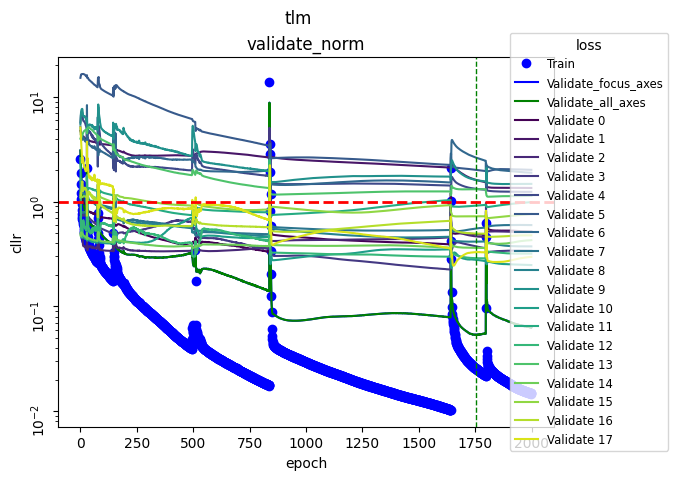}
    \caption{Loss plot of 'pairs TLM-formula' neural net}
    \label{fig:pair_1_within_var_double_lossplot}
\end{figure}

Figure \ref{fig:pair_1_within_var_double_lossplot} shows the loss plot for the standard TLM NN ('Pairs TLM-formula', double projection in \ref{fig:grand_summary}). Dark blue circles show the train Cllr, while a dark green line shows the validation Cllr over epochs. Also, since we de-correlated the 18-dimensional space, we can also include Cllrs for the LRs per dimension (0 to 17). De red horizontal dashed line denotes a Cllr of 1 (neutral performance) while a green dashed vertical line denotes the epoch for which the validation loss was smallest (and these parameter values were retained). 

First, it can be seen that the loss-plot shows some sharp peaks, it does not decrease continuously. This is probably due to instability in the QR decomposition. We will investigate this further in the future. For now, it seems that this spiking is not detrimental, we are able to fit models despite the spiking.

The model at best-epoch (1754) has a Cllr of 0.053, and this is the median value for the 21 runs. At this point, most axes have a Cllr below one, but there are some axes that have Cllr above 1, i.e. they individually perform worse than neutral. As we will see for the other models below, this observation (that there are some axes that individually perform worse than neutral) is consistent over models. However, this does not mean that they contribute negatively to the models overall LRs. 

We found that this is one way the models compensate for poor fit of the normal-assumptions to the data. We tried calculating LRs using only axes which Cllr $\leq$ 1 (that's why the 'Validate-focus-axes' line is in the plot), but it was found that this degrades overall performance. It is speculated that the 'large-Cllr-axes' are used to 'fix' LLRs for a few detrimental comparisons (that cause large deviations from normality), making them much better while giving in a little on the bulk. The balance for the whole dataset is that the Cllr decreases.

\begin{figure}[H]
    \centering
    \includegraphics[width=1.2\textwidth]{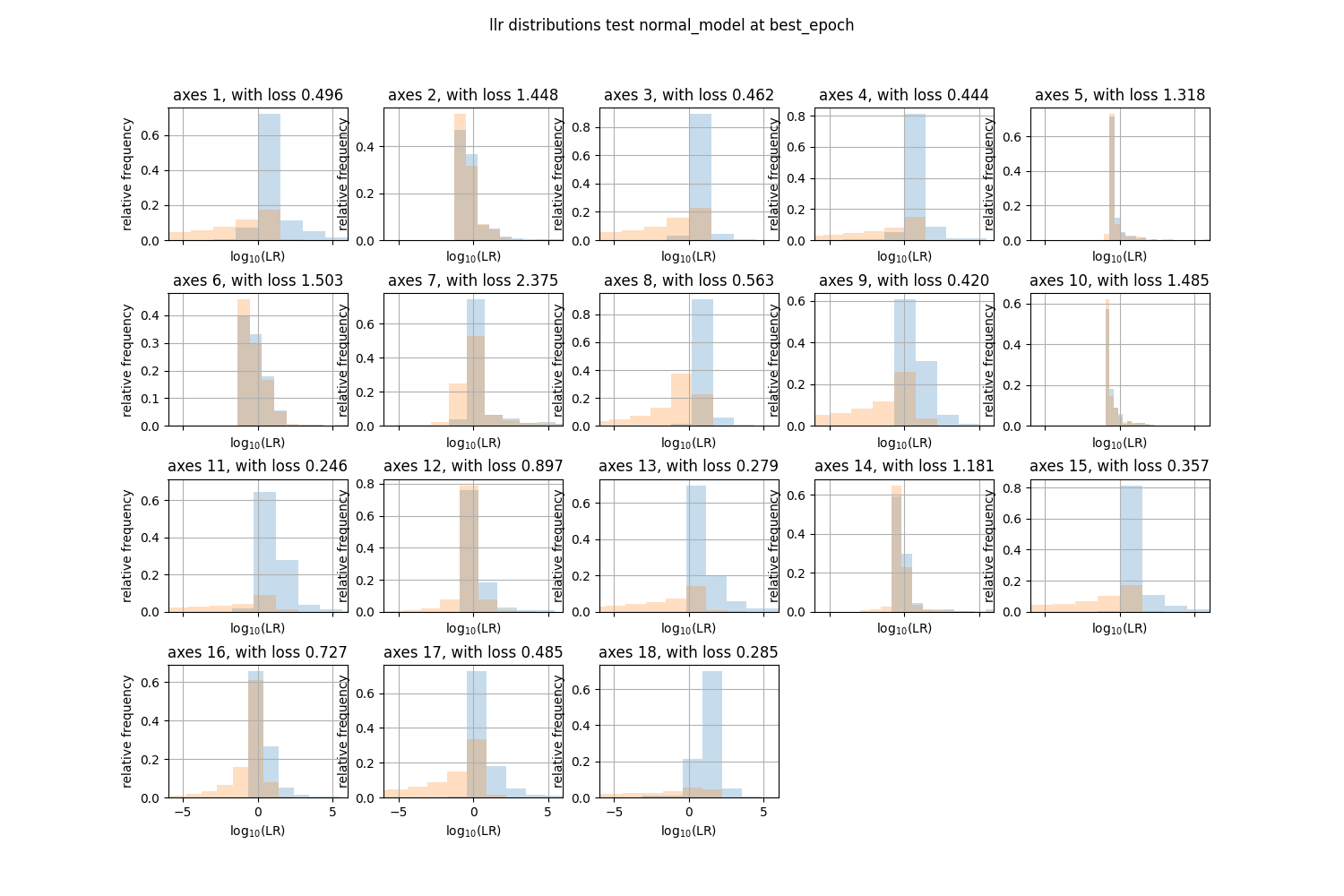}
    \caption{LLR-histograms of 'pairs TLM-formula' neural net. The loss denotes the Cllr on the test data for this axis.}
    \label{fig:pair_1_within_var_double_llrhistograms}
\end{figure}

Figure \ref{fig:pair_1_within_var_double_llrhistograms} shows the LLR histograms (orange = different-source, blue = same-source) for the individual axes. Above each histogram, the Cllr for this axis is depicted. Discrimination and Cllr vary widely over axes. Axes 1 has a relatively small Cllr of 0.496 and accordingly shows a relatively good separation between the two histograms. Axes 2 on the other hand, has a relatively large Cllr of 1.448 and accordingly shows bad separation between the two histograms. For some axes (2, 6) the two histograms are almost identical but LLRs still vary. An obvious improvement may be that for these axes, the model puts all LLRs at 0. Apparently, the model is not willing to do so under the current specification.

\begin{figure}[H]
    \centering
    \includegraphics[width=1\textwidth]{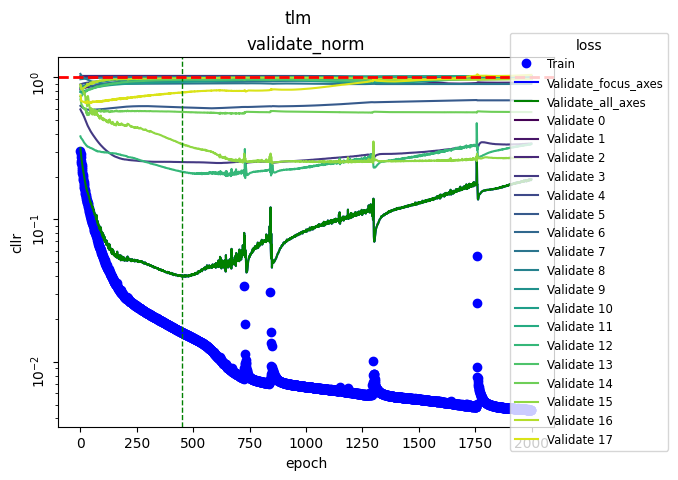}
    \caption{Loss plot of TLM-formula neural net with common projection matrix for between and within covariance.}
    \label{fig:pair_1_within_var_single_lossplot}
\end{figure}

Figure \ref{fig:pair_1_within_var_single_lossplot} shows the loss plot for the median the NTLM-model with the constraint that the projection matrix for the within and between covariance is the same ('Pairs TLM-formula', single projection in Fig. \ref{fig:grand_summary}). There are clear differences between this loss plot and its 'double projection' counterpart (\ref{fig:pair_1_within_var_double_lossplot}). The train loss still shows some, but fewer, spikes and it decreases more consistently. Furthermore, there are some axes with Cllr $\leq$ 1, but they are consistently more close to 1 than there 'double-projection' counterparts.

\begin{figure}[H]
    \centering
    \includegraphics[width=1.2\textwidth]{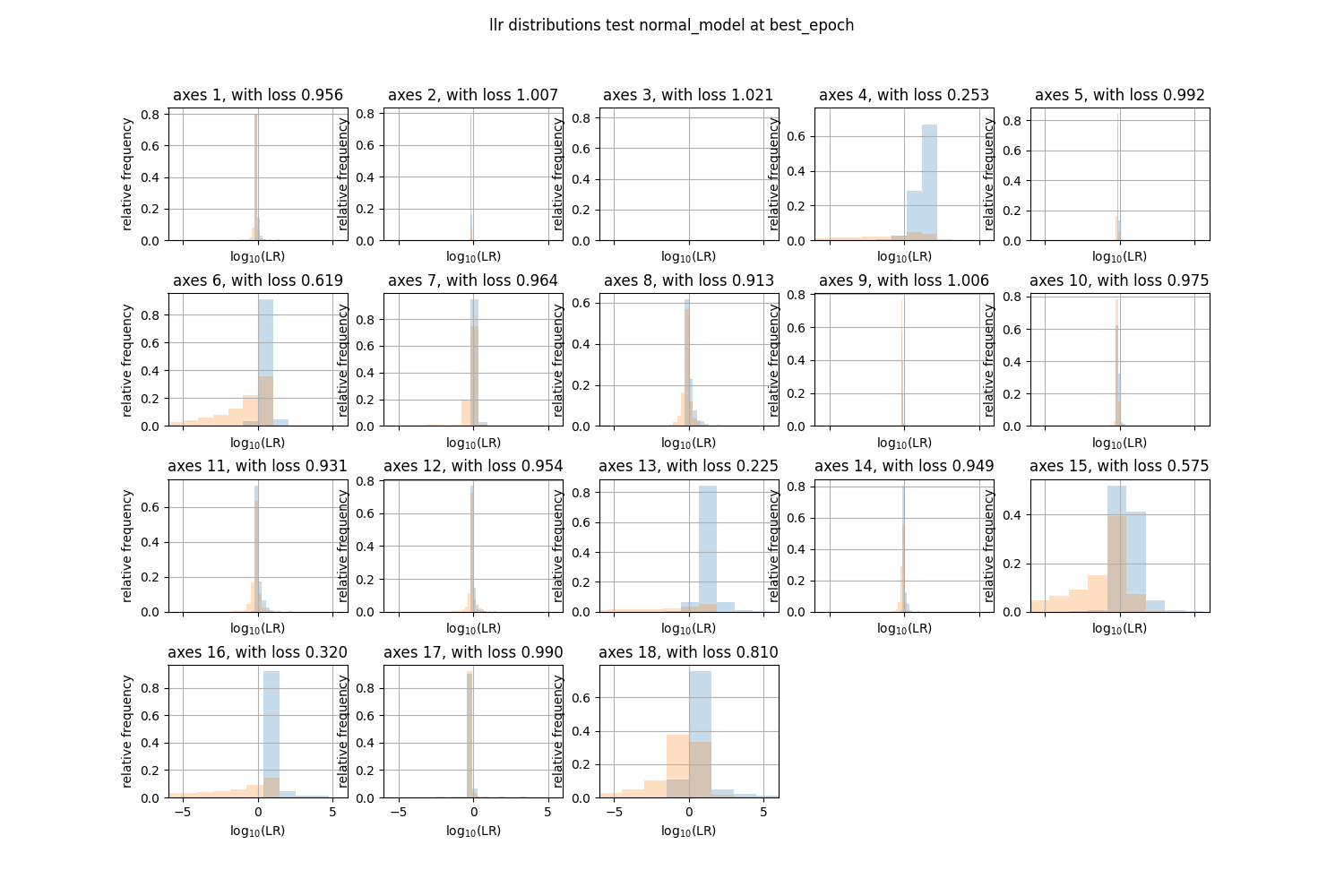}
    \caption{LLR-histograms of TLM-formula neural net with common projection matrix for between and within covariance.}
    \label{fig:pair_1_within_var_single_llrhistograms}
\end{figure}

Figure \ref{fig:pair_1_within_var_single_llrhistograms} shows the accompanying LLR-histograms for the test data. We see a new phenomena. For axes 1, 2, 3, 5, 9, 10, 12, 14, and 17, both histograms show a spike at LLR = 0, and there Cllr is almost 1. Apparently, this model configuration finds a way to exclude 'bad' axes by making them 'non-contributing'. We also studied the values of the variances, and the model does this by 'blowing up' the within and between variance, making them much larger than the data mean-squared-deviations.

\begin{figure}[H]
    \centering
    \includegraphics[width=1\textwidth]{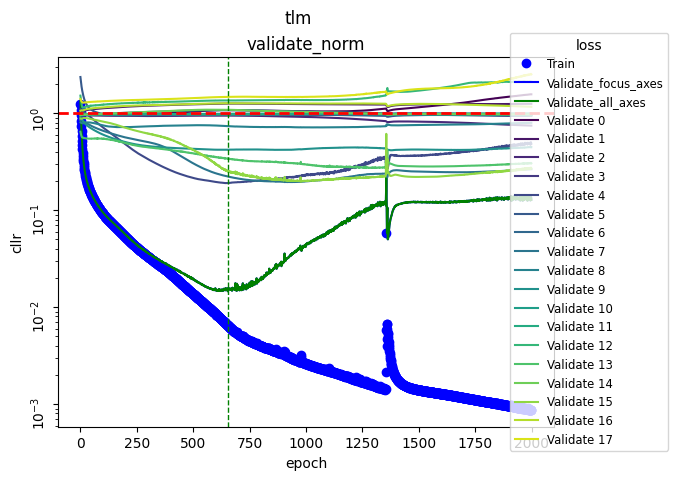}
    \caption{Loss plot of TLM-formula neural net with common projection matrix for between and within covariance, and different variance vectors for CSE and SRE.}
    \label{fig:pair_2_within_var_single_lossplot}
\end{figure}

Figure \ref{fig:pair_2_within_var_single_lossplot} shows the loss plot where the CSE and the SRE variance-vectors of the NTLM are allowed to be different (and with a common projection matrix), see model 'Pairs. N=w, Single' in Fig. \ref{fig:grand_summary}). This plot shows that 'bad axes' have Cllrs that are distinctively larger than 1 again. This may hint at worse performance but the contrary is true: as shown in \ref{fig:grand_summary} this model has better average performance than the 'same within-variance (TLM-formula)' counterpart.

\begin{figure}[H]
    \centering
    \includegraphics[width=1.2\textwidth]{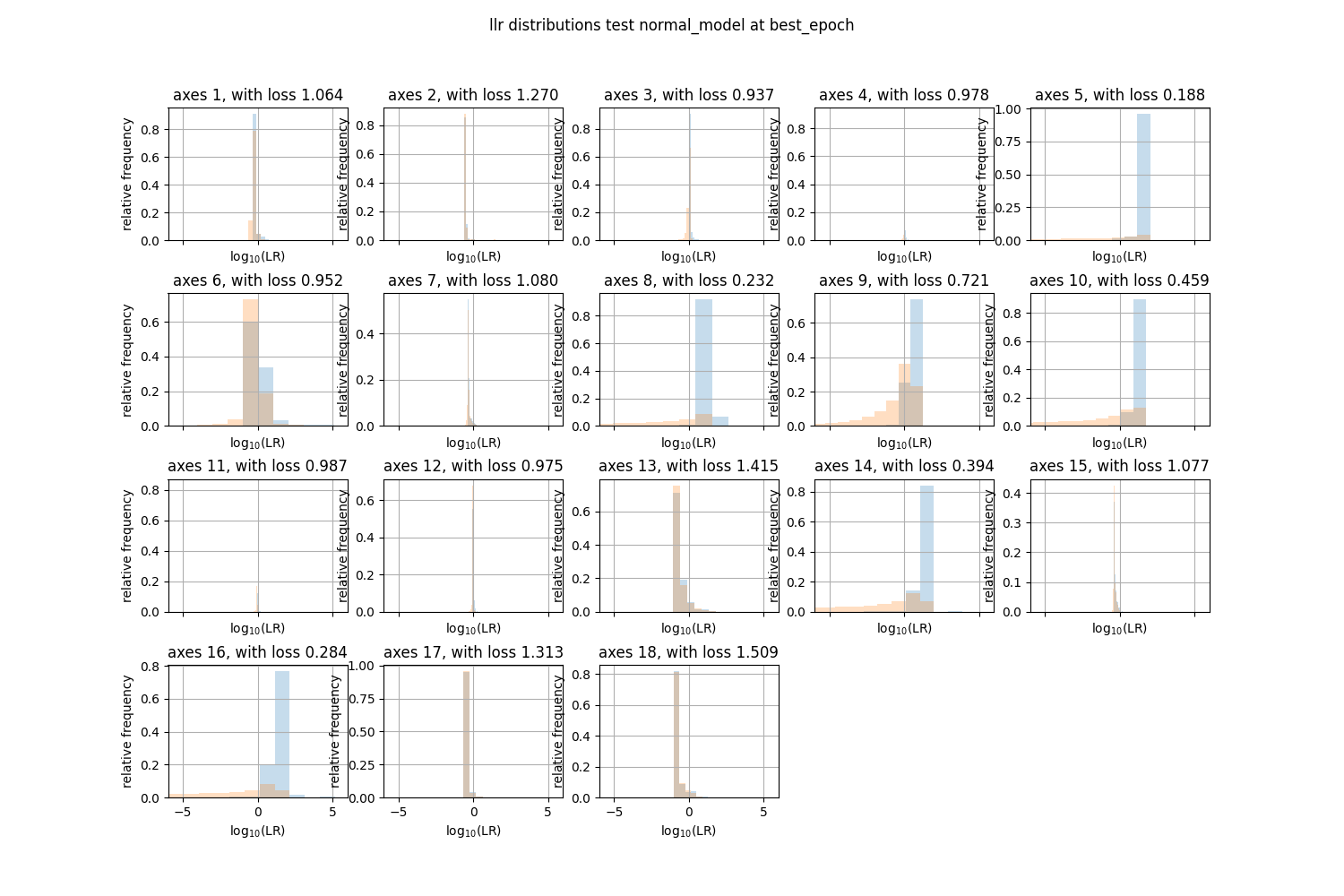}
    \caption{LLR-histograms of TLM-formula neural net with common projection matrix for between and within covariance, and different variance vectors for CSE and SRE.}
    \label{fig:pair_2_within_var_single_llrhistograms}
\end{figure}

Figure \ref{fig:pair_2_within_var_single_llrhistograms} shows what this 'within-variance-freedom' does to the LLR-histograms per axis. Again, several axes show histogram spikes (axes 2, 3, 4, 7, 11, 12, 15), but some of them are not centered at LR = 1 anymore. Axes 2, 7 and 15 introduce a 'bias' to the LR-system, shifting LLRs to more negative values. For same-source LLRs, this bias is compensated for by axes 5, 8, 9, 10, 14 and 16. The overall balance is a performance gain as shown in \ref{fig:grand_summary}.

Next, we show more detailed results for our best performing model, 'Source, N=2, varying', with one projection matrix.

\begin{figure}[H]

\begin{subfigure}{0.3\textwidth}
\includegraphics[width=1\linewidth]{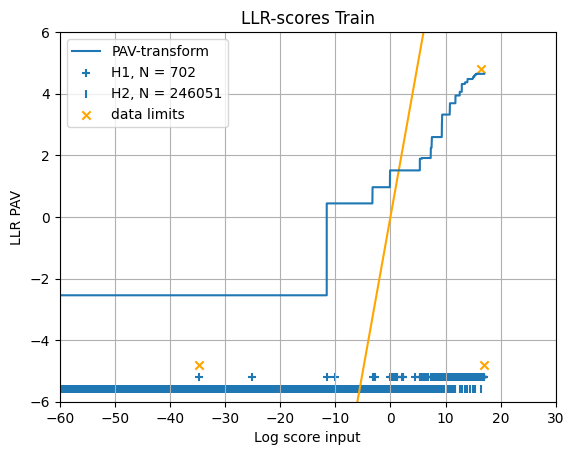} 
\caption{Uncalibrated TLM}
\label{fig:calibration_raw_TLM}
\end{subfigure}
\begin{subfigure}{0.3\textwidth}
\includegraphics[width=1\linewidth]{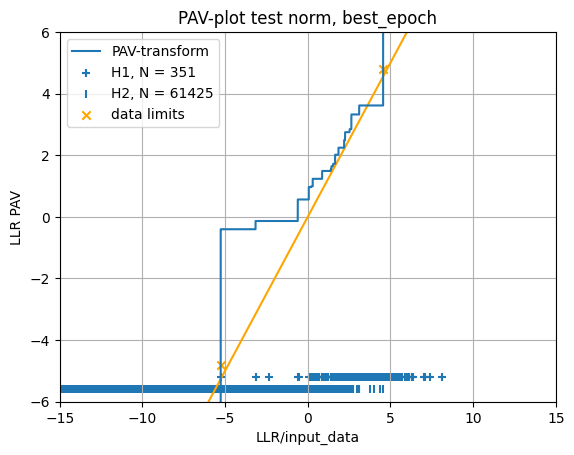}
\caption{Normal predictions}
\label{fig:calibration_normal}
\end{subfigure}
\begin{subfigure}{0.3\textwidth}
\includegraphics[width=1\linewidth]{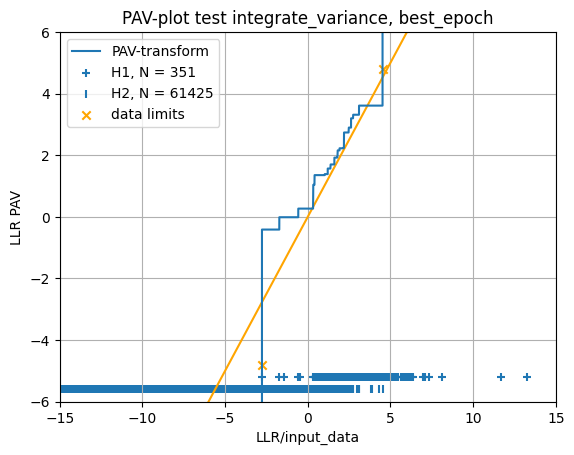}
\caption{$\sigma_w$ Integrated out}
\label{fig:calibration_t}
\end{subfigure}

\caption{Calibration plots for test data showing how close the PAV-operation to the LRs is to the yellow reference identity line. a) Typical calibration plot for TLM output. b) Typical calibration plot for a 'Sources, N=2' single model. c) Typical calibration plot for a 'Sources, N=2, varying' single model. At the bottom of the figures, $H_2$ and $H_1$ LLRs of the test data are plotted as '$+$'and '$|$' respectively.}
\label{fig:calibration_plots_compared}
\end{figure}

Figure \ref{fig:calibration_plots_compared} compares a typical calibration plot of the best model (Sources, N=2, varying, single projection) in panel c to typical calibration plots of uncalibrated ML-estimated TLM output (panel a) and the former model but without within-variance integration in panel b. The closer the blue line is to the yellow line, the better the calibration. 

From panel a it is clear why raw TLM-output needs post-hoc calibrating. The output is far from well-calibrated LRs, with a few very small false negative LLRs on the left, to generally positive-biased LLRs in the middle and right range. The middle and right plot (both NTLM models) show much better calibration. For the middle plot, still some small false negative LLRs are observed, but not as detrimental as for the uncalibrated TLM-system. The mid- and right range of the plot are well-calibrated. The plot in panel \ref{fig:calibration_t} shows improvement of the calibration of false negatives as compared to Fig. \ref{fig:calibration_normal}, which is the result of integrating out the within-variance uncertainty. Due to relaxing the 'common within variance' assumption, the contraction of false negatives is generally more strong as compared to integrating out the variance assuming a common within variance over sources (not shown). This is due to the retainment of more uncertainty in the variance-distribution when a non-equal within-variance over sources is assumed.

\begin{figure}[H]

\begin{subfigure}{0.3\textwidth}
\includegraphics[width=1\linewidth]{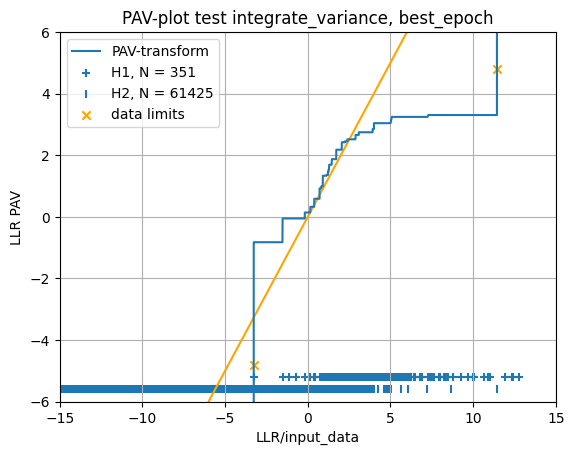} 
\caption{Worst Cllr-validate (0.068), Cllr-test = 0.063}
\label{fig:best_model_worst}
\end{subfigure}
\begin{subfigure}{0.3\textwidth}
\includegraphics[width=1\linewidth]{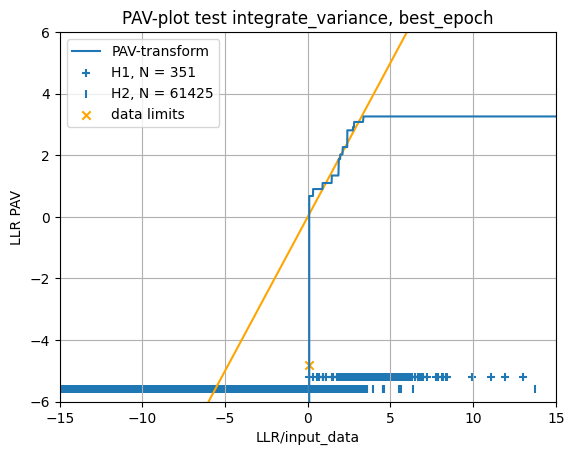}
\caption{Median Cllr-validate (0.025), Cllr-test = 0.022}
\label{fig:best_model_median}
\end{subfigure}
\begin{subfigure}{0.3\textwidth}
\includegraphics[width=1\linewidth]{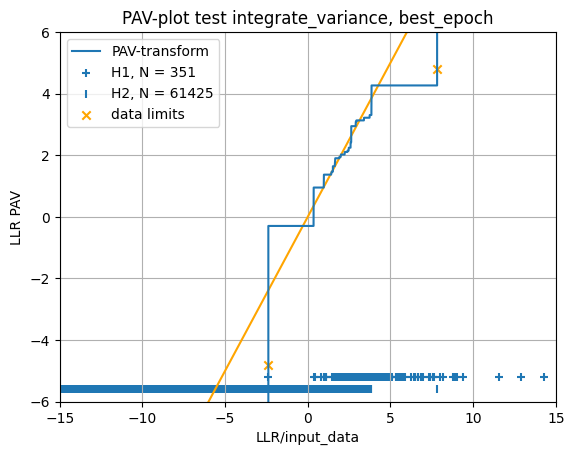}
\caption{Best Cllr-validate (0.012), Cll-test = 0.023}
\label{fig:best_model_best}
\end{subfigure}
\caption{Calibration plots on test data for the best model ('Sources, N=2, varying, single'). a) worst run on validation Cllr, b) median run on validation Cllr, c) best run on validation Cllr.}
\label{fig:calibration_plots_runs}
\end{figure}

To get a feeling for how well the best model is calibrated over the 21 runs, figure \ref{fig:calibration_plots_runs} shows calibration plots (on the test data) for the worst, median and best runs, sorted on Cllr-validate. For plot a, the worst case, calibration is off for small and for large LLRs while the middle range is well-calibrated. Plot b, the median case, is overall well-calibrated except for strong positive LLRs, that are strongly miscalibrated due to a few false-positive LLR-values. Plot c, the best case, shows a generally well-calibrated curve, except that LLRs at the extremes are somewhat too strong, roughly one order of magnitude. For all runs, calibration is clearly better than using raw maximum likelihood TLM LLRs as in Figure \ref{fig:calibration_raw_TLM}.

\begin{figure}[H]
    \centering
    \includegraphics[width=1.2\textwidth]{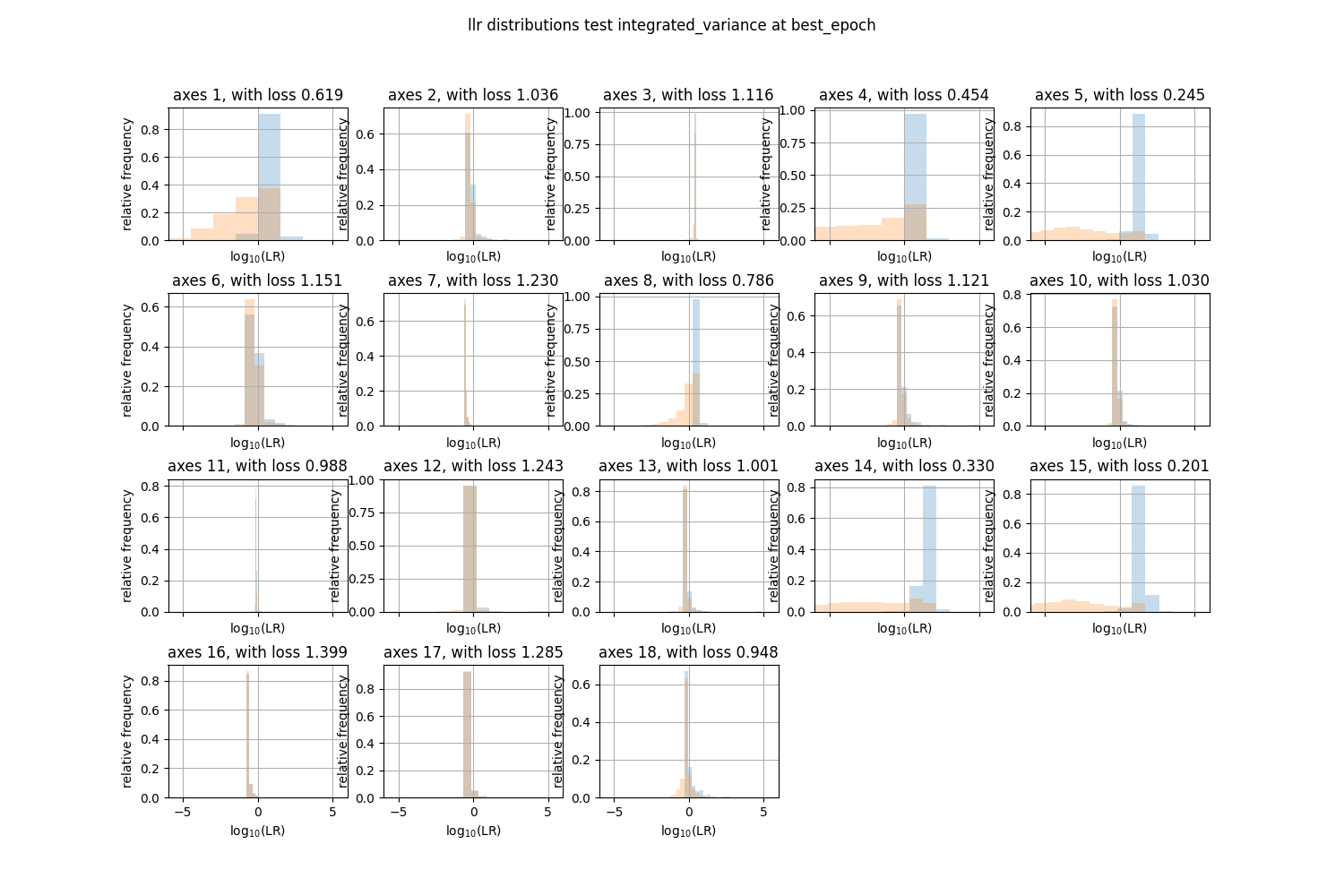}
    \caption{LLR histograms for the median run (Cllr on validation set) of the best model.}
    \label{fig:llr_histograms_best_model}
\end{figure}

Figure \ref{fig:llr_histograms_best_model} shows the LLR-histograms per axis for the best-model's median run. Qualitatively, the plot does not look very different from the LLR-histograms in Figure \ref{fig:pair_1_within_var_single_llrhistograms} (the 'Pairs, N=2, single' variant). The plot shows 'bias' for axes 3, 7 and 16, and discriminating axes 1, 4, 5, 14 and 15.

\begin{figure}[H]
    \centering
    \includegraphics[width=1.2\textwidth]{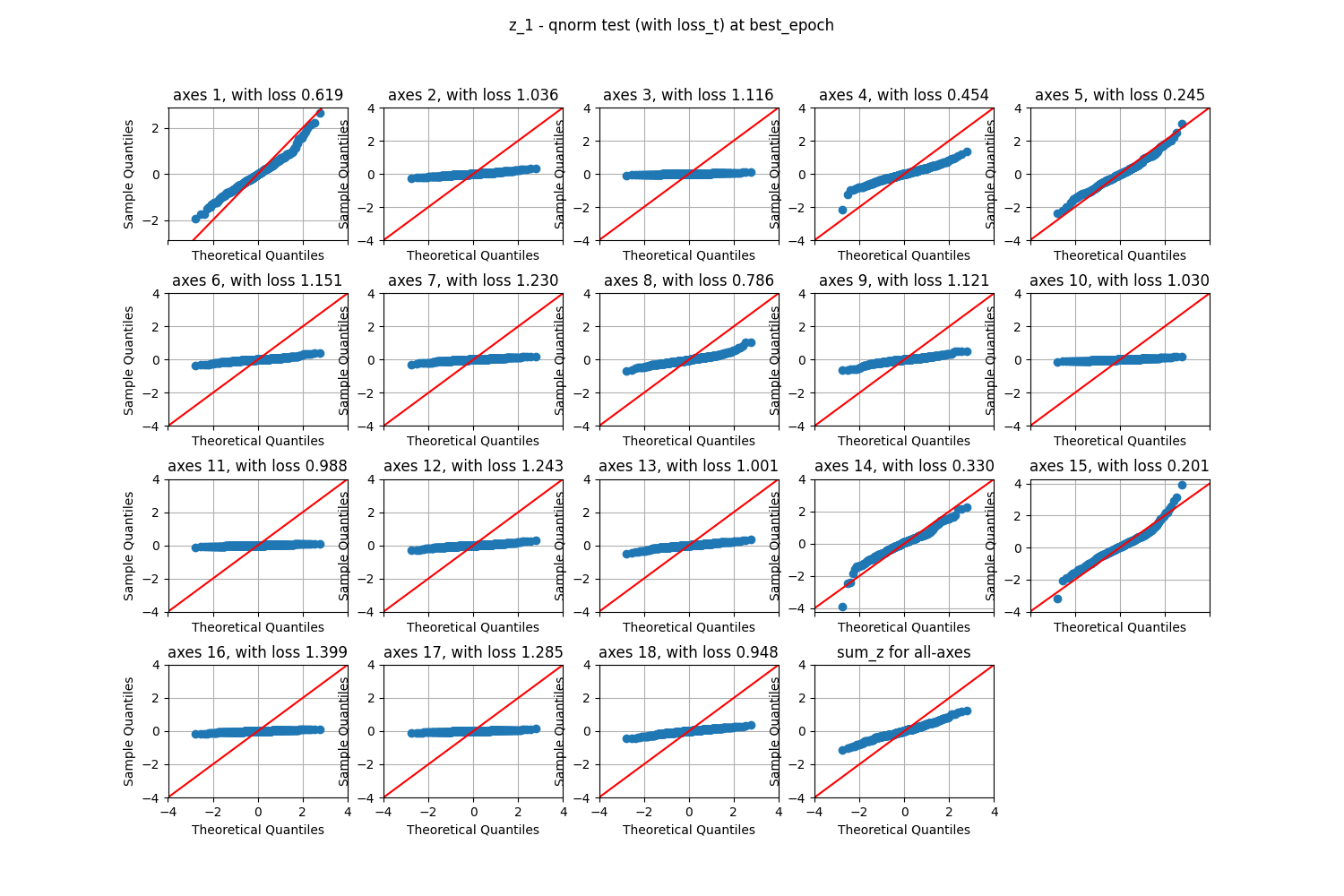}
    \caption{QQnorm-plots for the median run (Cllr on validation set) for 'z-scores' same-source test data under the $H_1$ model.}
    \label{fig:QQnorm-plots_h1_best_model}
\end{figure}

We can further study the behavior of these different axis-types by looking at QQnorm-plots per axis. Figure \ref{fig:QQnorm-plots_h1_best_model} shows QQnorm-plots for projected then scaled (PTS) same-source data under $H_1$. It does not make sense to look at QQnorm-plots for PTS different-source data since they are not supposed to follow the $H_1$-model. PTS same-source data is close to a standard normal distribution when their blue dots follow the red line. When blue dots follow a straight line with a smaller slope, PTS data is normally distributed with smaller data-variance than the variance-parameter fitted by the model. The plot shows that PTS data for the discriminating axes identified in figure \ref{fig:llr_histograms_best_model}: 1, 5, 14, 15 and more-or-less 4 follow a standard normal distribution in congruence with the model definition. The PTS data for all other axes have much smaller data-variance than fitted variance, leading to a more or less constant $H_1$ probability for all same-source data.

\begin{figure}[H]
    \centering
    \includegraphics[width=1.2\textwidth]{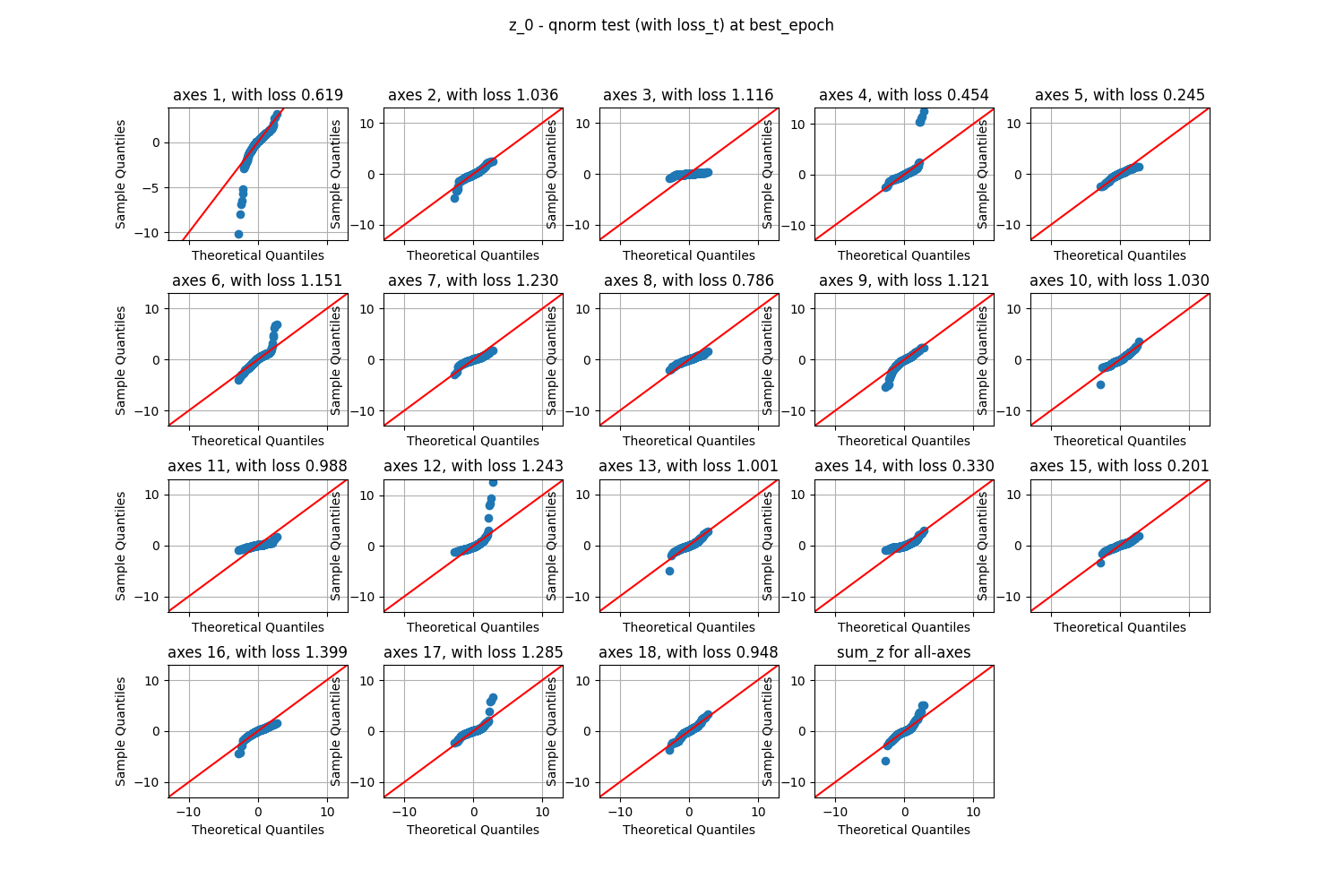}
    \caption{QQnorm-plots for the median run (Cllr on validation set) for 'z-scores' of same-source test data under the $H_2$ model.}
    \label{fig:QQnorm-plots_h2_best_model}
\end{figure}

 Figure \ref{fig:QQnorm-plots_h2_best_model} shows QQnorm-plots for PTS same-source data under $H_2$. The general trend is that this PTS data is roughly standard normally distributed for all axes, except for the tail-distributions, that can be quite deviant. For 'discriminating' axes identified in figure \ref{fig:llr_histograms_best_model} (axes 1, 4, 5, 14 and 15), the PTS data for axes 5, 14 and 15 follow a standard normal distribution closely, while for axes 1 and 4 the tails show outliers. Regarding the 'bias' axes 3, 7 and 16, the PTS data for axis 3 has a smaller variance than the fitted variance whereas axes 7 and 16 show more-or-less standard normally distributed data.
 
\begin{figure}[H]

\begin{subfigure}{0.45\textwidth}
\includegraphics[width=1\linewidth]{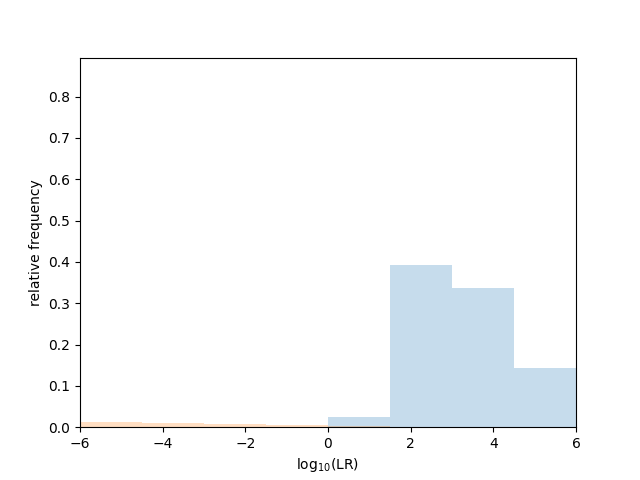} 
\caption{}
\label{fig:best_model_llr_histogram}
\end{subfigure}
\begin{subfigure}{0.45\textwidth}
\includegraphics[width=1\linewidth]{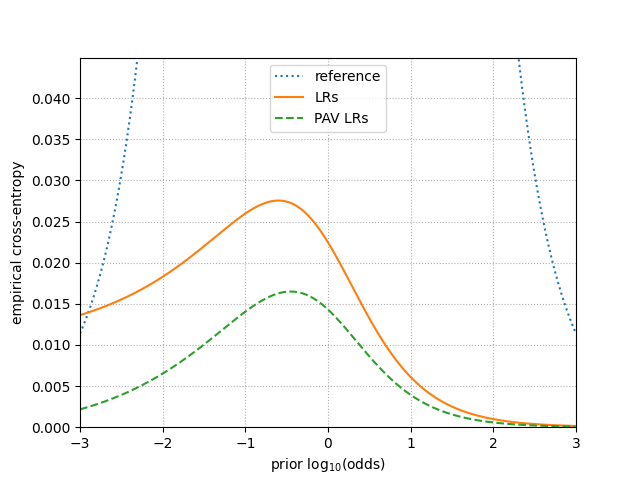}
\caption{}
\label{fig:best_model_ece}
\end{subfigure}
\caption{Omnibus-LLR histogram (a) and ECE-plot (b) for the median (Cllr on validation set) run, for test data.}
\label{fig:best_model_omnibus_llr_histogram_and_ece}
\end{figure}

Figure \ref{fig:best_model_omnibus_llr_histogram_and_ece} shows the LLR-histogram of the omnibus LLRs and the ECE-plot\cite{RamosEntropy} for the test data (median run for Cllr validate). Concerning figure \ref{fig:best_model_llr_histogram}, note that most of the different-source LLRs are off-scale having values much smaller than -6. The plot shows exceptionally good separation, as is usual for LA-ICP-MS measurements on glass fragments \cite{EsVergeerGlassLRs}. 

The ECE-plot in Fig. \ref{fig:best_model_ece} shows lines for a reference system (LR=1 always, blue dots), for the LR-system itself (orange) and for the LR-system after PAV-calibration on its LLR values (green dashes). It plots ECE as a function of the log prior odds. Where a perfect LR-system, that outputs LLRs = -infinite and LLRs = infinite for different-source and same-source LLRs respectively, produces an ECE-value of 0 irrespective of the log prior odds, a neutral system produces an ECE-value of 1 at prior-odds of 1 as its maximum value\cite{RamosEntropy}. 

The smaller the ECE-values for the data and the LR-system the better. ECE for this LR-system (on test data) is better than an 'LR=1' system except for log prior odds $\leq$ -3. The latter unwanted result shows the importance of being well-calibrated; if the LR-system would be well-calibrated over the whole LR-range it would be better than the neurtral system irrespective of the prior odds. From Fig. \ref{fig:calibration_t} it can be seen that calibration is off for large LLR values, which causes this behavior of the ECE-line at small prior odds. The Cllr (ECE at log prior odds = 0) for this run is equal to 0.0225. Calibration error is usually measured at log prior odds = 0, by subtracting ECE from ECE-after-PAV\cite{RamosEntropy, RamirezGaussianization}. The calibration error is 0.007 (30$\%$), roughly a factor of two smaller than the best model in \cite{RamirezGaussianization} trained on comparable BKA glass data, for which the calibration error is about 70$\%$.

\section{Discussion, future plans and conclusion}

For our glass LA-ICP-MS dataset, we have shown that we can improve performance and calibration of neural-net TLM LR-systems leveraging the eigenvalue-eigenvector decomposition of covariance matrices in combination with Bayesian probability calculus. Our baseline 'raw' NTLM has roughly the same performance as the best 'classically' trained TLM-version with Gaussianization in \cite{RamirezGaussianization}, while our 'Source, N=2, varying within variance' model with one projection matrix improves performance (as measured by Cllr) by a factor of 4.5 and shows a strong improvement in calibration compared to a classically trained TLM. For glass data, as far as the authors are aware, these are the best feature-based LR-systems reported in the literature.
Using separate variances and projection matrices allowed for the formulation of the 'Source, N=2, varying within variance' NTLM in the form of a joint projection matrix for the within and between space and allowing the model to choose different values for the within-variance vectors for CSE and SRE. The space of orthonormal projection matrices was spanned using QR-decomposition.

However, we speculate that QR-decomposition is also responsible for the spikes in the loss plot. When two-projection models are trained, spikes are more frequent than when one-projection models are trained. It is speculated that minor updates in the random-matrix parameter values that QR-decomposition operates on, can lead to relatively large changes in the projection matrix.

The variance-orthonormal decomposition also allowed us to decorrelate the omnibus LLR and study properties per axis. This makes the behavior of our NTLMs more easy to study than counterparts in the literature. When fixing orthonormal matrices or extending the number of within variance vectors, we were able to study the change in fitting behavior in the internals of the TLM, using QQnorm plots and LLR-histograms per axis. We saw unexpected (at least for us) phenomena such as the occurrence of 'bias axes' and the expansion of the $H_1$ variance far beyond data-variance. We have already leveraged what we have learned from this analysis in our implementation that numerically integrates out the within-variance uncertainty: 'variance blow-up' was our motivation for replacing the data-variance by the sampled within-variance from our fitted prior distribution, and not the within-variance parameter itself.

What also was unexpected is the complex response to increasing or decreasing the number of trainable parameters. First, it was observed that a decrease of trainable parameters (by going to one projection matrix instead of two) improved performance. Later, it was observed that increasing the number of parameters (going from one to two within-variance vectors) improved performance further. Using one within-variance vector, the ratio $H_1$-variance versus $H_2$-variance is at most 2. Using two within-variance vectors, this ratio can be any natural number, and this additional freedom the model used to improve performance. 
\newline
Apparently, it matters in what direction one gives the model more freedom and in what direction one applies a restriction. Our findings were mostly based on trial-and-error and had little theoretical guidance. It would be nice if theoreticians could provide guidance and maybe lead us to new directions.

Integration of variance uncertainty also improved model performance. Two methods were used. The first one assumes the same variance for all sources, and integrated out the uncertainty of the total $H_1$ and $H_2$ variance using t-distributions. The second one assumed a varying within variance over sources and integrated out this uncertainty by numerical means finding an appropriate prior distribution describing this uncertainty. The last procedure was applied post-hoc using the validation data to calculate the sums-of-squares. It was found that the latter method gave the largest performance gain.

Post-hoc methods have the added advantage that numerical methods come into scope anyhow. If one would apply such a numerical method during training, it would have to be redone at the end of every epoch, and this would require a relatively large amount of computation time. 
Moreover, from a Bayesian perspective, it makes sense to have as objective during training an optimized single-valued parameter, while for out-of-sample prediction after training one integrates out the uncertainty of this parameter. 
\newline
We had the choice to either integrate out the uncertainty governed by the training data, or by the validation data. We choose to use the validation data, since this is more representative for out-of-sample data than the training data. Note that this post-hoc procedure would also facilitate fast training for 'Heavy-tailed NPLDA' in \cite{BurgetBrummerNPLDA}, training their fast NPLDA model and than fitting the Heavy-tailed PLDA model post-hoc.

We think that, for glass data, our results are encouraging for the performance of feature-based LR-systems. Feature-based LRs have the intrinsic advantage that similarity and typicality are retained as identifiable factors for later interpretation purposes. Post-hoc calibrated feature-based LRs do not have this property, while score-based LRs have been criticized\cite{MorrisonScoreBasedTypicality, MorrisonTakingAccountOfTypicality, NeumannDefenceAgainstTwo} for not taking into account typicality at all. Feature-based LRs may therefore be preferable from an interpretation point of view.
However, from a Bayesian decision theoretic point of view, one is indifferent to how one arrives at a number to update the prior odds. As long as the system to produce these numbers is well-calibrated for a relevant dataset, any method goes\cite{VergeerRankingTheStars, BoonstraReconciling}. On the other hand, it is clear that in principle, feature-based LR-systems capture more information and they have the potential to perform better than score-based\cite{VergeerRankingTheStars} and post-hoc calibrated feature-based LR-systems. Nevertheless, reality is stubborn. Our best neural feature-based TLM still performs a factor of 1.5 (on Cllr) worse than our classically trained state-of-the-art post-hoc calibrated TLM model. 

However, we believe there is ample opportunity for further improvement.
\begin{itemize}
\item First, it is desirable to remove the spikes in the loss plots. It is hypothesized that the QR-decomposition operation is responsible for the spikes. We will try to replace this operation by other procedures that span the space of orthonormal matrices.
\item Second, the between space is clearly not multivariate normally distributed, see Fig. \ref{fig:BetweenBivariateScatter}. In order to have a better model for the between distribution, we plan to replace the normal model for the between distribution by a KDE model.
\item Third, overfitting mainly manifests itself in false-negative outliers (and also to a lesser extent false-positive outliers). Integrating out the within variance has a diminishing effect on the false negatives.  We plan to include further measures that remedy the occurence of outliers.
\item Fourth, in the literature complementary methods are explored, using Gaussianizing\cite{RamirezGaussianization, RibeiroEffectsGausianization} transformations that bring the data to the model. It would be interesting to implement these in combination with our NTLMs.
\item Fifth, one could leverage the finetuning ability of gradient descent learning. One could first pretrain a NTLM on other, datasets (for this work for example the ones in \cite{RamirezGaussianization}), and then finetune for on the local situation using local data. This may remedy overfitting as well.
\end{itemize}

Apart from trying to improve the NTLM for glass data, there are other paths that may be worth exploring as well. One of them is the application of our models to other datasets and other data modalities, such as firearm toolmarks image data\cite{BaikerToolmarkGun} or speaker data\cite{RibeiroEffectsGausianization}. For these data types, one can either train a separate 'preprocessor', possibly a deep neural net, that brings the data to the model, and then train the NTLM in a separate training run, or go for end-to-end training, optimizing all model parameters in a joint gradient descent loop.

Concluding, we have shown for a forensic 'small data' glass dataset improvement in calibration and performance of feature-based 'two-level-model' LR-systems by training them as a neural network. Making modifications to the NTLM-model, we improve a factor of 4.5 as compared to a baseline NTLM-model, and also a factor of 4.5 on 'classically' trained Gausianized data measured under similar circumstances\cite{RamirezGaussianization}. Nevertheless, our benchmark state-of-the art post-hoc calibrated feature-based model still performs a factor of 1.5 better. For the current dataset, we have come a long way closing the gap between performance of feature-based models and the state-of-the-art post-hoc calibrated model, squeezing the gap in terms of Cllr from a factor of 6.7 to 1.5. In the future we will try to close the gap completely and aim to study generalization of our models on other data modalities. All-in-all, we think the results are encouraging and we hope that these spark more research in this field.

\bibliographystyle{unsrt}
\bibliography{bibtex.bib}
\end{document}